\documentclass{elsartL}
\usepackage{graphicx}
\usepackage{amssymb}
\newsavebox{\myhbar}
\savebox{\myhbar}{$\hbar$}
\usepackage{amsmath}
\usepackage{mathrsfs}
\usepackage{mathtools}
\usepackage{rotating}
\usepackage{appendix}
\begin{document}

\begin{frontmatter}
\title{Low-energy parameterisations of the pion-nucleon phase shifts}
\author{Evangelos Matsinos}
\begin{abstract}
Compared in this work are a few sets of results, obtained from the pion-nucleon ($\pi N$) data at low energy (pion laboratory kinetic energy up to $100$ MeV) on the basis of the modelling of the $s$- and $p$-wave $K$-matrix 
elements (or of their reciprocal) via simple polynomials. The fitted values and uncertainties of the model parameters, as well as the corresponding Hessian (covariance) matrices, for three of these parameterisations are 
given in tabular form for two types of joint fits: to the measurements of the two elastic-scattering processes $\pi^\pm p \to \pi^\pm p$, and to those of the $\pi^+ p$ reaction and of the $\pi^- p$ charge-exchange reaction 
$\pi^- p \to \pi^0 n$. From these results, reliable and (largely) data-driven (hence model-independent) predictions, accompanied by uncertainties which reflect the statistical and systematic fluctuation of the input data, 
can be obtained for the low-energy constants of the $\pi N$ interaction (scattering lengths/volumes and range parameters), for the phase shifts, for the $K$-matrix elements, and for the partial-wave amplitudes in the 
(dominant at low energy) $s$ and $p$ waves. After the addition of the $d$- and $f$-wave contributions, and the inclusion of the electromagnetic effects, corresponding predictions can be obtained for the usual low-energy 
$\pi N$ observables, i.e., for the differential cross section and for the analysing power.
\end{abstract}
\end{frontmatter}

\section{\label{sec:Introduction}Introduction}

I intended to write a paper on the low-energy parameterisations of the pion-nucleon ($\pi N$) $K$-matrix elements (equivalently, of the $\pi N$ phase shifts) for some time, but more urgent matters got in the way and I had 
to postpone this task until better times. I am somewhat relieved that the time is now ripe for me to get back on track and accomplish this task.

I discovered the importance of these model-independent ways of accounting for the $\pi N$ phase shifts after Fettes and I sought in 1995 \cite{Fettes1997} the description of the $\pi^+ p$ measurements at low energy (for 
pion laboratory kinetic energy $T \leq 100$ MeV) in simpler ways than calling on the use of a newly-introduced (at that time), and considerably more complicated, $\pi N$ interaction model \cite{Goudsmit1994} based on hadronic 
exchanges. In fact, the very idea of developing a model-independent way of analysing the low-energy $\pi N$ data emanated from the requirement to put that hadronic model to the test.

On the basis of the information, which can be obtained from the tables of this work, as well as from the uploaded ancillary material, reliable and (largely) data-driven predictions, accompanied by uncertainties which 
reflect the statistical and systematic fluctuation of the input data, can be obtained for the low-energy constants of the $\pi N$ interaction (scattering lengths/volumes and range parameters), for the phase shifts, for the 
$K$-matrix elements, and for the partial-wave amplitudes in the (dominant at low energy) $s$ and $p$ waves. After the addition of the $d$- and $f$-wave contributions and the inclusion of the electromagnetic (EM) effects, 
corresponding predictions can be obtained for the observables of the three $\pi N$ reactions which are experimentally accessible at low energy, namely of the two elastic-scattering (ES) processes $\pi^\pm p \to \pi^\pm p$ 
and of the $\pi^- p$ charge-exchange (CX) reaction $\pi^- p \to \pi^0 n$.

Of relevance in this work is the modelling of the $K$-matrix elements (or of the reciprocal of these quantities) using simple polynomials. The low-energy expressions, developed and/or used within the context of the 
Chiral-Perturbation Theory ($\chi$PT), will not be discussed. Also not discussed are more elaborate expressions, developed in order to account for the $\pi N$ phase shifts in a broad energy domain, e.g., see Ref.~\cite{Gibbs2005}. 
This study is structured as follows. After attending to a few definitions in the beginning of the subsequent section, four schemes for modelling the $\pi N$ phase shifts will be detailed. The results of their application to 
the low-energy $\pi N$ measurements are given in Section \ref{sec:Results}, which is divided into four parts. Section \ref{sec:Procedure} provides information about the general procedure which is followed in new analyses 
performed within the context of the ETH $\pi N$ project: this procedure is divided into two phases, which in turn are further subdivided into steps. Used in the modelling of the $s$- and $p$-wave $K$-matrix elements in the 
first phase, which is relevant in this work, is any of the four aforementioned parameterisations. Section \ref{sec:ResultsCommonDB} discusses the effectiveness of these parameterisations to account for the input data, common 
to all methods (for the purposes of a fair comparison) and devoid of outliers for any of these methods. From this comparison, one of the schemes for modelling the $\pi N$ phase shifts will not be pursued further. The outcome 
of the application of the remaining three schemes is discussed in Section \ref{sec:ResultsDifferentDB}. Starting from identical databases (DBs), outliers are removed separately for each method, as that method requires. All 
important results from that part of the study, i.e., the fitted values and uncertainties of the fourteen model parameters, as well as the Hessian (covariance) matrices from two types of fits, will become available for use, 
in the form of tables at the end of this study, as well as of one Excel file, uploaded as ancillary material. One interesting application of this study is discussed in Section \ref{sec:Application}. A summary of this work 
is given in the last section, Section \ref{sec:Conclusions}.

All rest masses and $3$-momenta will be expressed in energy units. Excepting the masses and partial decay widths of the four higher baryon resonances (HBRs) discussed in Section \ref{sec:Resonant}, the values of the physical 
constants were obtained from the 2020 compilation of the Particle-Data Group \cite{PDG2020}.

\section{\label{sec:Parameterisations}Four schemes for modelling the $\pi N$ phase shifts}

Before commencing the description of the schemes for modelling the $\pi N$ phase shifts at low energy, I will introduce the $S$-, $T$-, and $K$-matrix elements for the spin-isospin channels applicable to the $\pi N$ 
interaction. Each of these quantities is characterised by the value of the total isospin $I$, either $3/2$ or $1/2$. The orbital angular momentum of the $\pi N$ system couples with the nucleon spin in two ways (for $l>0$), 
resulting in a total angular momentum of either $l+1/2$ or $l-1/2$. The spin-isospin notation for the elements of a matrix $A$ follows $A^I_{l\pm}$, where the sign in the subscript identifies the total angular momentum for 
the given ($s$, $p$, $d$, $f$, \dots) orbital ($l=0$, $1$, $2$, $3$, \dots, respectively). The $S$-matrix element is linked to the corresponding phase shift $\delta$ via the expression:
\begin{equation} \label{eq:EQ001}
S^I_{l\pm} \coloneqq \exp \left( 2 i \delta^I_{l\pm} \right) \, \, \, ,
\end{equation}
where $i$ stands for the imaginary unit. The $T$-matrix element (partial-wave scattering amplitude) is related to the $S$-matrix element according to the expression:
\begin{equation} \label{eq:EQ002}
S^I_{l\pm} \coloneqq 1 + 2 i q T^I_{l\pm} \, \, \, ,
\end{equation}
where $q$ is the magnitude of the $3$-momentum vector in the centre-of-mass (CM) coordinate system: $q \coloneqq \lvert \vec{q} \, \rvert$. Evidently,
\begin{equation} \label{eq:EQ003}
T^I_{l\pm} = \frac{\exp \left( 2 i \delta^I_{l\pm} \right) - 1}{2 i q} = \frac{\sin \delta^I_{l\pm} \, \cos \delta^I_{l\pm}}{q} + i \frac{\sin^2 \delta^I_{l\pm}}{q} \, \, \, .
\end{equation}
Another convenient quantity in the modelling is the $K$-matrix element, introduced via the relation:
\begin{equation} \label{eq:EQ004}
K^I_{l\pm} \coloneqq \frac{1}{i q} \frac{S^I_{l\pm} - 1}{S^I_{l\pm} + 1} = \frac{\tan \delta^I_{l\pm}}{q} \, \, \, .
\end{equation}
In comparison with the literature (in particular, of the distant past), the notations in Eqs.~(\ref{eq:EQ002}-\ref{eq:EQ004}) may differ; for instance, the quantities $K^I_{l\pm}$ and $T^I_{l\pm}$ were frequently defined as 
dimensionless.

From Eqs.~(\ref{eq:EQ003},\ref{eq:EQ004}), it follows that
\begin{equation} \label{eq:EQ005}
T^I_{l\pm} = \frac{K^I_{l\pm}}{1 - i q K^I_{l\pm}} = \frac{K^I_{l\pm} \left( 1 + i q K^I_{l\pm} \right)}{1 + q^2 \left( K^I_{l\pm} \right)^2} \Rightarrow \frac{\Im [T^I_{l\pm}]}{\Re [T^I_{l\pm}]} = q K^I_{l\pm} = \tan \delta^I_{l\pm} \, \, \, ,
\end{equation}
where the operators $\Re$ and $\Im$ return the real and the imaginary part of a complex number, respectively.

\subsection{\label{sec:Hoehler}The effective-range expansion}

In his book \cite{Hoehler1983}, H{\"o}hler introduced the effective-range expansion in Section 3.5.2 (`Expansions at the $s$-channel threshold') as
\begin{equation} \label{eq:EQ006}
q^{2l+1} \cot \delta^I_{l\pm} = (a^I_{l\pm})^{-1} + b^I_{l\pm} q^2 + \mathcal{O}(q^4) \, \, \, ,
\end{equation}
where the quantity $a^I_{l\pm}$ is the ``scattering length'' and $b^I_{l\pm}$ is half the so-called ``effective range,'' which appears in Ref.~\cite{Hoehler1983} as $r^I_{l\pm}$ (in fact, as $r_{l\pm}$, given that the 
superscript $I$ is implied in that section of H{\"o}hler's book). Under a more modern convention, the scattering lengths pertain to the $s$ waves ($a^{3/2}_{0+}$ and $a^{1/2}_{0+}$), whereas the quantities $a^{I}_{1\pm}$ 
are known as scattering volumes. The scattering lengths are expressed in a variety of units: in fm, in MeV$^{-1}$ or GeV$^{-1}$, or in units of the reciprocal of the charged-pion rest mass ($m_c^{-1}$); by analogy, the 
scattering volumes are given in fm$^3$, in MeV$^{-3}$ or GeV$^{-3}$, or in $m_c^{-3}$. Although one might refer to the constant terms in the parameterisation of the higher waves (e.g., $d$ and $f$) as scattering lengths, 
such terminology would `raise an eyebrow' nowadays.

Equations (\ref{eq:EQ006}) imply that
\begin{equation} \label{eq:EQ007}
\frac{\tan \delta^I_{l\pm}}{q^{2l+1}} = \frac{1}{(a^I_{l\pm})^{-1} + b^I_{l\pm} q^2 + \mathcal{O}(q^4)} \, \, \, , \text{or, using Eqs.~(\ref{eq:EQ004}),}
\end{equation}
\begin{equation} \label{eq:EQ008}
\frac{\tan \delta^I_{l\pm}}{q} \coloneqq K^I_{l\pm} = \frac{q^{2l}}{(a^I_{l\pm})^{-1} + b^I_{l\pm} q^2 + \mathcal{O}(q^4)} \, \, \, .
\end{equation}

For the (dominant at low energy) $s$ and $p$ waves, Eqs.~(\ref{eq:EQ008}) suggest that
\begin{equation} \label{eq:EQ009}
\frac{\tan \delta^I_{0+}}{q} = K^I_{0+} = \frac{1}{(a^I_{0+})^{-1} + b^I_{0+} q^2 + \mathcal{O}(q^4)} \, \, \, \text{and}
\end{equation}
\begin{equation} \label{eq:EQ010}
\frac{\tan \delta^I_{1\pm}}{q} = K^I_{1\pm} = \frac{q^2}{(a^I_{1\pm})^{-1} + b^I_{1\pm} q^2 + \mathcal{O}(q^4)} \, \, \, .
\end{equation}
Given the additional $q^2$ in the numerators of the right-hand side (rhs) of Eqs.~(\ref{eq:EQ010}), an overall parameterisation of the quantities $K^I_{0+}$ and $K^I_{1\pm}$ to $\mathcal{O}(q^4)$ would suggest the use of the 
following expressions in the modelling of the two $s$- and the four $p$-wave phase shifts~\footnote{Given their current uncertainties, the low-energy $\pi N$ data cannot (reliably) determine more than seven parameters per 
isospin channel. Even with seven parameters, the correlations are generally strong, especially so between the parameters which enter the modelling of the same partial wave.}:
\begin{equation} \label{eq:EQ011}
K^I_{0+} = \frac{1}{(a^I_{0+})^{-1} + b^I_{0+} q^2 + c^I_{0+} q^4} \, \, \, \text{and}
\end{equation}
\begin{equation} \label{eq:EQ012}
K^I_{1\pm} = \frac{q^2}{(a^I_{1\pm})^{-1} + b^I_{1\pm} q^2} \, \, \, ,
\end{equation}
where the seven parameters $a^I_{0+}$, $b^I_{0+}$, $c^I_{0+}$, $a^I_{1+}$, $b^I_{1+}$, $a^I_{1-}$, and $b^I_{1-}$ per isospin channel must be determined from the $\pi N$ data. H{\"o}hler mentions expressions obtained 
after the $K$-matrix elements are expanded in $q^2$ (e.g., see the second expression in Eqs.~(A.3.62) of Ref.~\cite{Hoehler1983}), yet I will not consider any such expansions in this section.

H{\"o}hler also mentions in passing another popular parameterisation for the $s$ waves, that of the real part of $T^I_{0+}$. From Eqs.~(\ref{eq:EQ005},\ref{eq:EQ011}), one obtains the relation:
\begin{equation} \label{eq:EQ014}
\Re [T^I_{0+}] = a^I_{0+} - (a^I_{0+})^2 \left( a^I_{0+} + b^I_{0+} \right) q^2 + \mathcal{O}(q^4) \, \, \, .
\end{equation}

\subsection{\label{sec:ELW}The ELW parameterisation}

As mentioned in the previous section, H{\"o}hler himself had proposed other parameterisations of the $K$-matrix elements, one of which suggested the expansion of the rhs of Eq.~(\ref{eq:EQ006}) in $q^2$ and, as a result, 
did not contain the algebraic fractions of the original form. I cannot recollect studies in which such parameterisations were put into application before Ref.~\cite{Ericson2004} appeared. In that paper, a simple polynomial 
parameterisation of the $K$-matrix elements in $q^2$ was employed in the $s$-wave modelling of the (physical) $\pi^- p$ ES channel: $K_{\rm cc}$ was parameterised as
\begin{equation} \label{eq:EQ015}
K_{\rm cc} = a_{\rm cc} + b_{\rm cc} q^2 + \mathcal{O}(q^4) \, \, \, .
\end{equation}
In the spirit of such polynomial parameterisations,
\begin{equation} \label{eq:EQ016}
K^I_{0+} = a^I_{0+} + b^I_{0+} q^2 + c^I_{0+} q^4 \, \, \, \text{and}
\end{equation}
\begin{equation} \label{eq:EQ017}
K^I_{1\pm} = q^2 \left( a^I_{1\pm} + b^I_{1\pm} q^2 \right) \, \, \, . 
\end{equation}
In this work, I will refer to the modelling, based on Eqs.~(\ref{eq:EQ016},\ref{eq:EQ017}), as `ELW parameterisation', despite the fact that it was first suggested by H{\"o}hler.

\subsection{\label{sec:ETH1}The parameterisation used in the ETH $\pi N$ project}

The parameterisation, used in the ETH $\pi N$ project, was established in the mid 1990s \cite{Fettes1997}. The basic difference to the parameterisations of Sections \ref{sec:Hoehler} and \ref{sec:ELW} is that the 
expansion parameter is not $q^2$, but the pion CM kinetic energy $\epsilon$. As the expansion of $\epsilon$ in terms of $q$ does not involve odd powers, i.e.,
\begin{equation} \label{eq:EQ019}
\epsilon = \frac{q^2}{2 m_c} - \frac{q^4}{8 m^3_c} + \frac{q^6}{16 m^5_c} - \mathcal{O}(q^8) \, \, \, ,
\end{equation}
the polynomial parameterisations of the $K$-matrix elements in $\epsilon$ are not incompatible with Eqs.~(\ref{eq:EQ006}).

There was one reason why in Ref.~\cite{Fettes1997} the authors set out to examine the possibility of an improved parameterisation (in comparison with the effective-range expansion) of the $K$-matrix elements. The 
parameterisations of Eqs.~(\ref{eq:EQ011},\ref{eq:EQ012}), theoretically motivated as they might be, do not necessarily have to be optimal in terms of the description of the experimental data; one of the tasks in 
Ref.~\cite{Fettes1997} was to investigate that subject. Although the form of Eq.~(\ref{eq:EQ011}) provides some stability in the determination of the parameters entering $K^I_{0+}$, it was unclear whether or not the same 
complexity was called for in the $p$-wave part of the interaction, i.e., in Eqs.~(\ref{eq:EQ012}).

To cut a long story short, an expression of similar structure as Eq.~(\ref{eq:EQ011}) was used in Ref.~\cite{Fettes1997} in the $s$ waves:
\begin{equation} \label{eq:EQ020}
K^I_{0+} = \frac{1}{(a^I_{0+})^{-1} + b^I_{0+} \epsilon + c^I_{0+} \epsilon^2} \, \, \, ,
\end{equation}
whereas a simpler - in comparison with Eqs.~(\ref{eq:EQ012}) - parameterisation was introduced in the $p$ waves:
\begin{equation} \label{eq:EQ021}
K^I_{1\pm} = a^I_{1\pm} \epsilon + b^I_{1\pm} \epsilon^2 \, \, \, .
\end{equation}
In this work, I will refer to the modelling, based on Eqs.~(\ref{eq:EQ020},\ref{eq:EQ021}), as `ETH parameterisation'.

\subsection{\label{sec:ETH2}Yet another parameterisation}

The original idea in Ref.~\cite{Fettes1997} was to make use of the parameterisations of Eqs.~(\ref{eq:EQ011},\ref{eq:EQ012}), but consistently replace $q^2$ in the denominators of the algebraic fractions with $\epsilon$. 
The first attempts to describe the data involved the forms:
\begin{equation} \label{eq:EQ022}
K^I_{0+} = \frac{1}{(a^I_{0+})^{-1} + b^I_{0+} \epsilon + c^I_{0+} \epsilon^2} \, \, \, \text{and}
\end{equation}
\begin{equation} \label{eq:EQ023}
K^I_{1\pm} = \frac{q^2}{(a^I_{1\pm})^{-1} + b^I_{1\pm} \epsilon} \, \, \, .
\end{equation}
I included this parameterisation in the present study in order to demonstrate that, in comparison with the parameterisations of Sections \ref{sec:Hoehler}-\ref{sec:ETH1}, it provides a poorer description of the low-energy 
$\pi N$ data. Overall, Eqs.~(\ref{eq:EQ023}) are less successful than Eqs.~(\ref{eq:EQ021}) in the modelling of the $p$-wave part of the interaction.

\subsection{\label{sec:Resonant}Addition of the dominant contributions from nearby resonances}

Provided that the CM total energy $W$ of the $\pi N$ system remains well below the masses of any HBRs with (non-zero) branching fractions to $\pi N$ decay modes, the parameterisations of the $K$-matrix elements, detailed 
in Sections \ref{sec:Hoehler}-\ref{sec:ETH2}, are acceptable ways of modelling the $K$-matrix elements in the $s$ and $p$ waves, compatible with the low-energy behaviour of these quantities set forth by Eq.~(\ref{eq:EQ006}). 
However, one distinctive feature of the $\pi N$ interaction is that there are two such resonances with masses within reach of the energy domain in which the analyses, performed within the context of the ETH $\pi N$ project, 
are confined. These two states are:
\begin{itemize}
\item the $\Delta(1232)$ resonance, which affects $K^{3/2}_{1+}$ ($P_{33}$ partial wave), and
\item the $N(1440)$ resonance (also known as Roper resonance), which affects $K^{1/2}_{1-}$ ($P_{11}$ partial wave).
\end{itemize}
For the purposes of the modelling, it is therefore advisable to explicitly add the resonant contributions in the $P_{33}$ and $P_{11}$ partial waves. As the two aforementioned resonances will be taken into account, the 
contributions from two additional states (one in each of the two partial waves), with larger masses and sizeably smaller influence at low energy, may (and will) also be included.

Regarding the contribution from the $\Delta(1232)$ resonance, a singular (at $W=M_\Delta$) term is added to the `background' $K^{3/2}_{1+}$ term, obtained with any of the parameterisations of Sections \ref{sec:Hoehler}-\ref{sec:ETH2}. 
Also appending the effects of the $\Delta(1600)$ resonance, one ends up with the expression:
\begin{align} \label{eq:EQ024}
K^{3/2}_{1+} = \left( K^{3/2}_{1+} \right)_b & + \frac{\Gamma_\Delta M_\Delta}{2 q_\Delta^3 (p_{0 \Delta} + m_p)} \frac{(p_0 + m_p) q^2}{W (M_\Delta-W)}\nonumber\\
& + \text{Corresponding contribution from $\Delta(1600)$} \, \, \, ,
\end{align}
where
\begin{itemize}
\item $m_p$ and $p_0$ denote the rest mass and the CM total energy of the proton;
\item $M_\Delta$ and $\Gamma_\Delta$ stand for the Breit-Wigner mass and the partial decay width of the $\Delta(1232)$ resonance to $\pi N$ decay modes (the corresponding branching fraction is nearly $100~\%$); and
\item the quantities $q_\Delta$ and $p_{0 \Delta}$ represent the values of the variables $q$ and $p_0$ at $W=M_\Delta$.
\end{itemize}
The first term on the rhs of Eq.~(\ref{eq:EQ024}) is taken from Eqs.~(\ref{eq:EQ012},\ref{eq:EQ017},\ref{eq:EQ021},\ref{eq:EQ023}), according to the selected parameterisation, whereas the singular term has been obtained 
from Ref.~\cite{Matsinos2014}, see $K_{1+}$ in Eqs.~(39) and the corresponding $K^{3/2}_{1+}$ element (after the isospin decomposition of $K_{1+}$ is taken into account), as well as footnote 10 therein. The contribution 
from the $\Delta(1600)$ can be obtained from Section 3.4 of Ref.~\cite{Matsinos2014}, by simply replacing $M_\Delta$ and the partial decay width of the $\Delta(1232)$ resonance with the corresponding quantities of the 
$\Delta(1600)$ resonance.

Regarding the contribution from the Roper resonance (and the sizeably smaller contribution from the higher $P_{11}$ state),
\begin{equation} \label{eq:EQ025}
K^{1/2}_{1-} = \left( K^{1/2}_{1-} \right)_b + \sum_{i=1}^2 \frac{(\Gamma_N)_i (M_N)_i \left( (p_{0 N})_i + m_p \right)}{2 (q_N^3)_i \left( (M_N)_i+m_p \right)^2} \frac{(W + m_p)^2 q^2}{(p_0 + m_p) W \left( (M_N)_i - W \right)} \, \, \, ,
\end{equation}
where $(\Gamma_N)_i$ is the partial decay width of each contributing $P_{11}$ resonance to $\pi N$ decay modes and $(M_N)_i$ is the Breit-Wigner mass of that state. The quantities $q_N$ and $p_{0 N}$ denote the $q$ and $p_0$ 
values at the pole of each resonance ($W=(M_N)_i$). The singular terms in Eq.~(\ref{eq:EQ025}) were obtained from Ref.~\cite{Matsinos2014}, see Section 3.5.1 therein, in particular, Eq.~(54) for $K_{1-}$.

All physical properties of these resonances have been fixed from Ref.~\cite{Matsinos2020}; the corresponding uncertainties are not used. The contribution of the resonant part in Eq.~(\ref{eq:EQ024}) to $a^{3/2}_{1+}$ is 
about $36.0130$ GeV$^{-3}$ for the parameterisations of Sections \ref{sec:Hoehler}, \ref{sec:ELW}, and \ref{sec:ETH2}, whereas that of the resonant part in Eq.~(\ref{eq:EQ025}) to $a^{1/2}_{1-}$ (for the same 
parameterisations) is about $4.9616$ GeV$^{-3}$, see Table \ref{tab:Contributions} at the very end of this study; both numerical results must be multiplied by $2 m_c$ to yield the corresponding contributions in case of the 
ETH parameterisation. The contributions of the resonant parts to $b^{3/2}_{1+}$ and $b^{1/2}_{1-}$ are different for the parameterisations of Sections \ref{sec:Hoehler}-\ref{sec:ETH2}, see Appendix \ref{App:AppA} for details.

\section{\label{sec:Results}Results}

Despite the fact that the same symbols were used in the identification of the fourteen model parameters (seven per isospin channel) in the previous section, it ought to be borne in mind that these parameters are different 
(and, as a rule, have different dimensions) across the four parameterisations. Although the use of the same seven symbols to identify these quantities may be considered confusing by some, it nevertheless prevents a logistical 
nightmare of characters, which (moreover) would have to be looked up time after time by the reader.

After considering three last issues, we will be ready to attempt the description of the low-energy $\pi N$ measurements with the $s$- and $p$-wave $K$-matrix elements of Sections \ref{sec:Hoehler}-\ref{sec:ETH2}, suitably 
modified after the inclusion of the resonant contributions (see Section \ref{sec:Resonant}).

The first issue concerns the fixation of the $d$ and $f$ waves. The frequently-voiced opinion that ``such effects are small'' fails in kinematical regions in which these `small effects' are amplified because of the 
cancellation of the main contributions (destructive interference between the $s$- and $p$-wave contributions to the $\pi N$ scattering amplitude). Kinematical regions of such a cancellation include the backward angular 
region for the $\pi^- p$ ES reaction and the forward one for the $\pi^- p$ CX reaction, both at moderate energy, around $50$ MeV. (The former effects were experimentally explored in Ref.~\cite{Janousch1997}, the latter in 
Ref.~\cite{Fitzgerald1986}.) At these kinematical regions, the generally `small contributions' become important, if not dominant. Considering that the introduction of additional parameters, e.g., to account for the energy 
dependence of the $d$- and $f$-wave phase shifts, is beyond the bounds of possibility at present, these phase shifts must be fixed from external sources. In the analyses, performed within the context of the ETH $\pi N$ 
project, the $d$ and $f$ waves are imported from the SAID results, at present from their `current' phase-shift solution XP15 \cite{XP15}. As the SAID analyses make use of the high-energy $\pi N$ measurements, their estimates 
for the $d$- and $f$-wave phase shifts are expected to be reliable.

The second issue concerns the inclusion of the EM effects. All analyses performed within the context of the ETH $\pi N$ project during the past two decades have used the EM corrections developed at the University of Zurich 
\cite{Oades2007,Gashi2001a,Gashi2001b}. To my knowledge, this has been the only programme which addressed the issue of the EM corrections throughout the low-energy region (also including, in a consistent manner, the 
corrections which ought to apply to the measurements of the strong-interaction shift $\epsilon_{1 s}$ and of the total decay width $\Gamma_{1s}$ in pionic hydrogen). That effort culminated in the publication of the EM 
corrections for $\pi^+ p$ scattering \cite{Gashi2001a}, for $\pi^- p$ scattering \cite{Gashi2001b}, as well as for $\epsilon_{1 s}$ and $\Gamma_{1s}$ \cite{Oades2007}. The programme rested upon the determination of hadronic 
potentials which optimally account for the low-energy $\pi N$ DB in the scattering region, and made use of the same potentials to determine the corrections at the $\pi N$ threshold using a three-channel 
($\pi^- p \to \pi^- p, \pi^0 n, \gamma n$) calculation.

The last issue concerns the choice of the minimisation function, used in the optimisation of the description of the input data. The phase-shift analyses, performed within the context of the ETH $\pi N$ project (as well as 
those of the SAID group), make use of the minimisation function which Arndt and Roper introduced half a century ago \cite{Arndt1972}. All details can be found in Section 3.3 of a recent study \cite{Matsinos2022a}; there is 
no need to repeat them here.

\subsection{\label{sec:Procedure}General procedure in new analyses performed within the context of the ETH $\pi N$ project}

To facilitate the understanding of the procedure, which will be followed in the analysis of the low-energy $\pi N$ measurements in this study, I will first lay out the general method which is followed when new analyses are 
performed within the context of the ETH $\pi N$ project. The course of analysis comprises two phases; it has been standardised and largely automated since a long time. The first phase makes use of the parameterisations of 
the $K$-matrix elements. (Although any of the parameterisations of Sections \ref{sec:Hoehler}-\ref{sec:ETH2} may be selected in the user interface, only the ETH parameterisation of Section \ref{sec:ETH1} has been used in 
the various analyses up to this time.) The main task in this phase is the identification and the removal of the outliers from the low-energy $\pi N$ DB and (thus) the preparation of the input for the second phase of the 
analysis. The use of the parameterisations of the $K$-matrix elements for this task is important because they provide a model-independent way of identifying the outliers, one which is devoid of theoretical constraints (other 
than the expected low-energy behaviour of the $K$-matrix elements). In addition, measurements are marked as outliers on the basis of comparisons with the same type of data: for instance, a decision on whether or not a 
specific measurement of the $\pi^+ p$ DB is an outlier rests upon the assessment of its proximity to the \emph{bulk} of the low-energy $\pi^+ p$ data. This explains why it is important (within the context of the ETH $\pi N$ 
project) to develop ways of modelling the partial-wave amplitudes which can account for the low-energy $\pi N$ measurements in the best way possible.

The first phase of new analyses comprises six steps, each one involving a different input DB. At each step, a loop
\begin{equation*}
A \rightleftharpoons B \, \, \, ,
\end{equation*}
where
\begin{itemize}
\item[$A$] represents the operation `Fit to the DB' and
\item[$B$] represents the operation `Remove from the DB the most discrepant outlier in the fit',
\end{itemize}
is set until all outliers are removed from the DB which is treated at that step. 

At the end of each cycle (one optimisation run) of each step, the p-values of the description of the datasets~\footnote{Each p-value is calculated from the $\chi^2$ contribution and the number of the active degrees of 
freedom of each dataset in the DB, i.e., of the number of datapoints which currently comprise the dataset. The absolute normalisation of each dataset is also subjected to testing (and occasional removal, in which case the 
number of degrees of freedom (NDF) of the dataset is the current number of its datapoints reduced by one), see Ref.~\cite{Matsinos2017a} for details.}, which comprise the DB at that cycle, are compared in order that the 
worst-described dataset be identified. If the p-value, corresponding to the description of that dataset, is below a user-defined significance threshold $\mathrm{p}_{\rm min}$, then the worst-described entry of that dataset 
(largest contribution to the $\chi^2$ value of that dataset) is removed from the DB (one outlier at a time) and the fit to the updated DB (former DB without the newly-marked outlier) is performed. The loop is repeated until 
the p-values of all datasets in the DB (which is treated at that step) exceed $\mathrm{p}_{\rm min}$, in which case the analysis enters the next step.

A few words about the choice of the significance threshold $\mathrm{p}_{\rm min}$ are in order. In the analyses, which are performed within the context of the ETH $\pi N$ project, $\mathrm{p}_{\rm min}$ is chosen to correspond 
to the frequency of occurrence of $2.5 \sigma$ effects in the normal distribution. This $\mathrm{p}_{\rm min}$ value is approximately equal to $1.24 \cdot 10^{-2}$, i.e., slightly exceeding $1.00 \cdot 10^{-2}$, which is 
the threshold regarded by most statisticians as the outset of statistical significance. (To ensure the consistency of new analyses, the entire procedure is routinely repeated for $\mathrm{p}_{\rm min}$ values associated 
with $2$ and $3 \sigma$ effects in the normal distribution.)

After these explanations, it is time I described the six steps of the first phase and the three steps of the second phase in every new analysis; they are as follows.
\begin{enumerate}
\item Separate fits to the $\pi^+ p$ DB (starting from the initial low-energy $\pi^+ p$ DB, as it has been detailed in Ref.~\cite{Matsinos2017a}) by variation of the three (one $s$-wave and two $p$-wave) $I=3/2$ partial-wave 
amplitudes (seven parameters in total). The $I=3/2$ partial-wave amplitudes are fixed from the final fit.
\item Separate fits to the $\pi^- p$ ES DB (starting from the initial low-energy $\pi^- p$ ES DB \cite{Matsinos2017a}) by variation of the three $I=1/2$ partial-wave amplitudes (seven parameters in total). The final $I=3/2$ 
partial-wave amplitudes of step (1) are used (central values, no uncertainties).
\item Separate fits to the $\pi^- p$ CX DB (starting from the initial low-energy $\pi^- p$ CX DB \cite{Matsinos2017a}) by variation of the three $I=1/2$ partial-wave amplitudes (seven parameters in total). The final $I=3/2$ 
partial-wave amplitudes of step (1) are (again) used (central values, no uncertainties).
\item Joint fits to the DB of the two ES reactions by variation of all six (two $s$-wave and four $p$-wave) $I=3/2$ and $I=1/2$ partial-wave amplitudes (fourteen parameters in total). Any additional outliers from this step 
are removed only when the DBs of both ES reactions are submitted to the optimisation at a later time; in the majority of the cases, these fits yield no further outliers.
\item Joint fits to the DB of the $\pi^+ p$ and of the $\pi^- p$ CX reactions by variation of all six $I=3/2$ and $I=1/2$ partial-wave amplitudes (fourteen parameters in total). Any additional outliers from this step are 
removed only when the DBs of both these reactions are submitted to the optimisation at a later time; in the majority of the cases, these fits yield no further outliers.
\item Global fits to all three DBs by variation of all six $I=3/2$ and $I=1/2$ partial-wave amplitudes (fourteen parameters in total). Any additional outliers from this step are removed only when global fits to all data 
are performed at a later time. Such fits are carried out only for the sake of completeness; the results have never been reported and/or used.
\item Joint fits to the DB of the two ES reactions using the ETH model \cite{Matsinos2017a} (seven parameters: coupling constants and admixture parameters). There is no identification of outliers at this step.
\item Joint fits of the DB of the $\pi^+ p$ and of the $\pi^- p$ CX reactions using the ETH model \cite{Matsinos2017a} (seven parameters). There is no identification of outliers at this step.
\item Global fits to all three DBs using the ETH model \cite{Matsinos2017a} (seven parameters). There is no identification of outliers at this step. Such fits are carried out only for the sake of completeness; the results 
have never been reported and/or used.
\end{enumerate}
The DBs are only forwards-updated (no backward feedback): the outliers from the separate fits of steps (1)-(3) are permanently removed from the corresponding DBs at all next steps, whereas those from the joint fits of steps 
(4) and (5) are removed only in case of the same-type joint fits or of global fits to all data. Relevant in this work are only steps (1)-(5) above.

\subsection{\label{sec:ResultsCommonDB}Results of the application of the parameterisations of Sections \ref{sec:Hoehler}-\ref{sec:ETH2} to the same data}

Starting from the initial low-energy $\pi N$ DB, as it has been detailed in Ref.~\cite{Matsinos2017a}, the parameterisations of Sections \ref{sec:Hoehler}-\ref{sec:ETH2}, modified according to Section \ref{sec:Resonant} and 
including the $d$- and $f$-wave contributions, as well as the EM effects (as previously described), were successively applied to the data, in the order they were introduced in this study. Following the steps (1)-(5) of the 
previous section, each input DB was updated; this procedure resulted in the creation of five `clean' DBs, i.e., data containing no outliers for \emph{any} of the parameterisation methods of Sections \ref{sec:Hoehler}-\ref{sec:ETH2}. 
Consequently, the parameterisations of Sections \ref{sec:Hoehler}-\ref{sec:ETH2} will be treated on an equal footing in the ensuing comparison, and the results of the fits to these five `clean' DBs may be taken as indicative 
of the effectiveness of each parameterisation in capturing the low-energy behaviour of the $\pi N$ partial-wave amplitudes.

The differences in the results of the fits to the five `clean' low-energy $\pi N$ DBs, listed in Table \ref{tab:ResultsCommonDB}, are not striking. On the other hand, striking differences are hardly expected given the 
affinity between of the parameterisation schemes of this study. Nevertheless, one result stands out when comparing the final $\chi^2$ values: the application of the parameterisation of Section \ref{sec:ETH2} results in the 
largest $\chi^2$ values in four of the five cases. Evidently, that parameterisation provides a poorer modelling of the $p$-wave part of the scattering amplitude of the two $\pi^- p$ (ES and CX) reactions. A similar, albeit 
less pronounced, effect can be seen in the description of the measurements in terms of the effective-range expansion (Section \ref{sec:Hoehler}): in comparison with the ELW and ETH parameterisations, the effective-range 
expansion is slightly more successful in the description of the $\pi^+ p$ data, but provides a poorer description of the measurements of the two $\pi^- p$ reactions. In the subsequent sections, results will be given for the 
parameterisations of Sections \ref{sec:Hoehler}-\ref{sec:ETH1}. The parameterisation of Section \ref{sec:ETH2} will not be pursued further.

\vspace{0.5cm}
\begin{table}[h!]
{\bf \caption{\label{tab:ResultsCommonDB}}}The $\chi^2$ values of the description of the five `clean' DBs of the first column using the parameterisations of Sections \ref{sec:Hoehler}-\ref{sec:ETH2}. NDF stands for the 
number of degrees of freedom in each case. 
\vspace{0.25cm}
\begin{center}
\begin{tabular}{|l|c|c|c|c|c|}
\hline
DB & NDF & Effective-range & ELW & ETH & Parameterisation of\\
 & & expansion \cite{Hoehler1983} & \cite{Ericson2004} & \cite{Fettes1997} & Section \ref{sec:ETH2}\\
\hline
\hline
$\pi^+ p$ & $414$ & $544.48$ & $550.92$ & $546.84$ & $546.41$\\
$\pi^- p$ ES & $322$ & $368.33$ & $346.73$ & $346.26$ & $375.67$\\
$\pi^- p$ CX & $322$ & $322.70$ & $308.35$ & $314.63$ & $324.78$\\
ES & $718$ & $859.94$ & $852.53$ & $846.25$ & $867.39$\\
$\pi^+ p$ and $\pi^- p$ CX & $736$ & $851.01$ & $842.13$ & $844.92$ & $854.92$\\
\hline
\end{tabular}
\end{center}
\vspace{0.5cm}
\end{table}

\subsection{\label{sec:ResultsDifferentDB}Results of the application of the parameterisations of Sections \ref{sec:Hoehler}-\ref{sec:ETH1} separately to the low-energy $\pi N$ data}

Steps (1)-(5) of Section \ref{sec:Procedure} were followed for each of the parameterisations of Sections \ref{sec:Hoehler}-\ref{sec:ETH1} separately: the application of each method resulted in the creation of sets of 
outliers, associated with that specific method. Initial and final $\chi^2$ values from the application of the three methods to the data are given in Tables \ref{tab:HOEHLER}, \ref{tab:ELW}, and \ref{tab:ETH}. Inspection 
of these tables suggests that the methods achieve comparable descriptions of the input data. Noticeable is only a marginal difficulty of the forms, associated with the effective-range expansion, to account for the $p$-wave 
part of the two $\pi^- p$ reactions.

The fitted values and uncertainties of the model parameters for the three parameterisations are given in Tables \ref{tab:HOEHLERParameters}, \ref{tab:ELWParameters}, and \ref{tab:ETHParameters} for two types of fit:
\begin{itemize}
\item joint fit to the measurements of the two ES reactions (joint fit A henceforth) and
\item joint fit to the measurements of the $\pi^+ p$ and of the $\pi^- p$ CX reactions (joint fit B henceforth).
\end{itemize}
In both cases, the isospin $I=3/2$ phases shifts are - to a great extent - determined from the $\pi^+ p$ reaction, leaving the determination of the $I=1/2$ phase shifts to the corresponding $\pi^- p$ (ES or CX) reaction. 
The Hessian matrices from these fits are given in Tables
\begin{itemize}
\item \ref{tab:HOEHLERHessian1} and \ref{tab:HOEHLERHessian2} for the effective-range expansion,
\item \ref{tab:ELWHessian1} and \ref{tab:ELWHessian2} for the ELW parameterisation, and
\item \ref{tab:ETHHessian1} and \ref{tab:ETHHessian2} for the ETH parameterisation.
\end{itemize}
These results are also uploaded as ancillary material, in the form of an Excel file; more precise information (i.e., more decimal places) about the fitted values and uncertainties of the model parameters can be obtained 
from that file.

The application of the parameterisations of the effective-range expansion (Section \ref{sec:Hoehler}) brought about one unexpected result. At step (4) of the procedure described in Section \ref{sec:Procedure}, one additional 
outlier was identified in one of the JORAM95 datasets, namely in the $\pi^+ p$ dataset at $44.60$ MeV \cite{Joram1995}; the identification of additional outliers at step (4) does not occur frequently in new analyses. The new 
outlier was one measurement in the Coulomb peak, corresponding to CM scattering angle $\theta=14.26^\circ$ ($(d\sigma/d\Omega)_{\rm CM}=3.931(262)$ mb/sr). With the identification of this measurement as an outlier, the 
aforementioned dataset exceeded the allowed number of outliers (see Ref.~\cite{Matsinos2017b}, pp.~7-8) and had to be removed from the $\pi^+ p$ DB. This removal accounts for the difference in the NDF of the final fit to 
the low-energy ES data when using the parameterisation of Section \ref{sec:Hoehler} (e.g., compare relevant entries in Tables \ref{tab:HOEHLER}, \ref{tab:ELW}, and \ref{tab:ETH}). \emph{Dura lex sed lex}.

From these results, reliable and (largely) data-driven predictions, accompanied by uncertainties which reflect the statistical and systematic fluctuation of the input data, can be obtained for a variety of physical 
quantities using any of the parameterisations of Sections \ref{sec:Hoehler}-\ref{sec:ETH1}:
\begin{itemize}
\item for the low-energy constants of the $\pi N$ interaction (scattering lengths/volumes and range parameters),
\item for the $\pi N$ phase shifts,
\item for the $K$-matrix elements, and
\item for the partial-wave amplitudes
\end{itemize}
in the (dominant at low energy) $s$ and $p$ waves. Last but not least, after the addition of the $d$- and $f$-wave contributions and the inclusion of the EM effects, corresponding predictions can be obtained for the 
observables of the three $\pi N$ reactions which can be subjected to experimental exploration at low energy. Regarding such predictions, one remark is in order. Due to the non-fulfilment of the triangle identity, see Eq.~(2) 
of Ref.~\cite{Matsinos2022a}, by the scattering amplitudes of the three low-energy $\pi N$ reactions, this work recommends the use of the results from the joint fit A for predictions associated with one of the ES processes, 
and from the joint fit B for predictions involving the $\pi^- p$ CX reaction.

\subsection{\label{sec:Application}One interesting application}

In this section, I will demonstrate how this work can be useful in the investigation of the violation of the isospin invariance in the $\pi N$ interaction at low energy. Let us assume that the objective is to compare the 
real parts of two $s$-wave $\pi^- p$ CX amplitudes: one of these amplitudes represents a prediction, based on the results of the fits to the ES data and making use of the so-called triangle identity (e.g., see Eq.~(2) of 
Ref.~\cite{Matsinos2022a}), whereas the other is obtained from fits involving the $\pi^- p$ CX data; in Ref.~\cite{Matsinos2022a}, the former amplitude is denoted by $f_{\rm CX}$, the latter by $f^{\rm extr}_{\rm CX}$. For 
the sake of variety, let me also use the results obtained from the ETH parameterisation.

In order that the two amplitudes be evaluated, needed as input are the fitted values and uncertainties of the fourteen model parameters, as well as the Hessian matrices from the joint fits A and B. This information will be 
retrieved from the tables of this work as follows:
\begin{itemize}
\item The optimal parameter values from the joint fits A and B will be obtained from Table \ref{tab:ETHParameters}. The fitted uncertainties, listed in the table, have already been corrected for the goodness of each fit via 
the application of the Birge factor $\sqrt{\chi^2/{\rm NDF}}$ \cite{Birge1932}, hence they are ready for use without further adjustment. As a result, $1 \sigma$ uncertainties will be obtained in the predictions, reflecting 
the statistical and systematic fluctuation of the data used in the two fits.
\item The Hessian matrix from the joint fit A will be obtained from Table \ref{tab:ETHHessian1}.
\item The Hessian matrix from the joint fit B will be obtained from Table \ref{tab:ETHHessian2}.
\end{itemize}

The $\pi^- p$ CX amplitude is constructed from the two isospin amplitudes according to the expression:
\begin{equation} \label{eq:EQ026}
T_{CX} = \frac{\sqrt{2}}{3} \left( T^{3/2} - T^{1/2} \right) \, \, \, ,
\end{equation}
implying that
\begin{equation} \label{eq:EQ027}
\Re [ T_{CX} ] = \frac{\sqrt{2}}{3} \left( \Re [ T^{3/2} ] - \Re [ T^{1/2} ] \right) \, \, \, .
\end{equation}
Given that the $s$-wave part of this amplitude reads as
\begin{equation} \label{eq:EQ027}
\Re [ T_{CX} ]_s = \frac{\sqrt{2}}{3} \left( \Re [ T^{3/2}_{0+} ] - \Re [ T^{1/2}_{0+} ] \right) \, \, \, ,
\end{equation}
the use of Eq.~(\ref{eq:EQ005}) leads to the expression:
\begin{equation} \label{eq:EQ028}
\Re [ T_{CX} ]_s = \frac{\sqrt{2}}{3} \left( \frac{K^{3/2}_{0+}}{1+q^2 \left( K^{3/2}_{0+} \right) ^2} - \frac{K^{1/2}_{0+}}{1+q^2 \left( K^{1/2}_{0+} \right)^2 } \right) \, \, \, .
\end{equation}
Of interest is the evaluation of the quantity $\Re [ T_{CX} ]_s$ from the results of the fits A and B. In each case, Monte-Carlo events will be generated, taking account of the fitted values of six model parameters ($a^I_{0+}$, 
$b^I_{0+}$, and $c^I_{0+}$), as well as of the two Hessian matrices. The generation of (single-precision) correlated random numbers in normal distribution is nearly effortless when using the standard CERN software library 
(functions CORSET and CORGEN). A short FORTRAN program, providing a solution to the problem of this section, is supplied as ancillary material.

\begin{figure}
\begin{center}
\includegraphics [width=15.5cm] {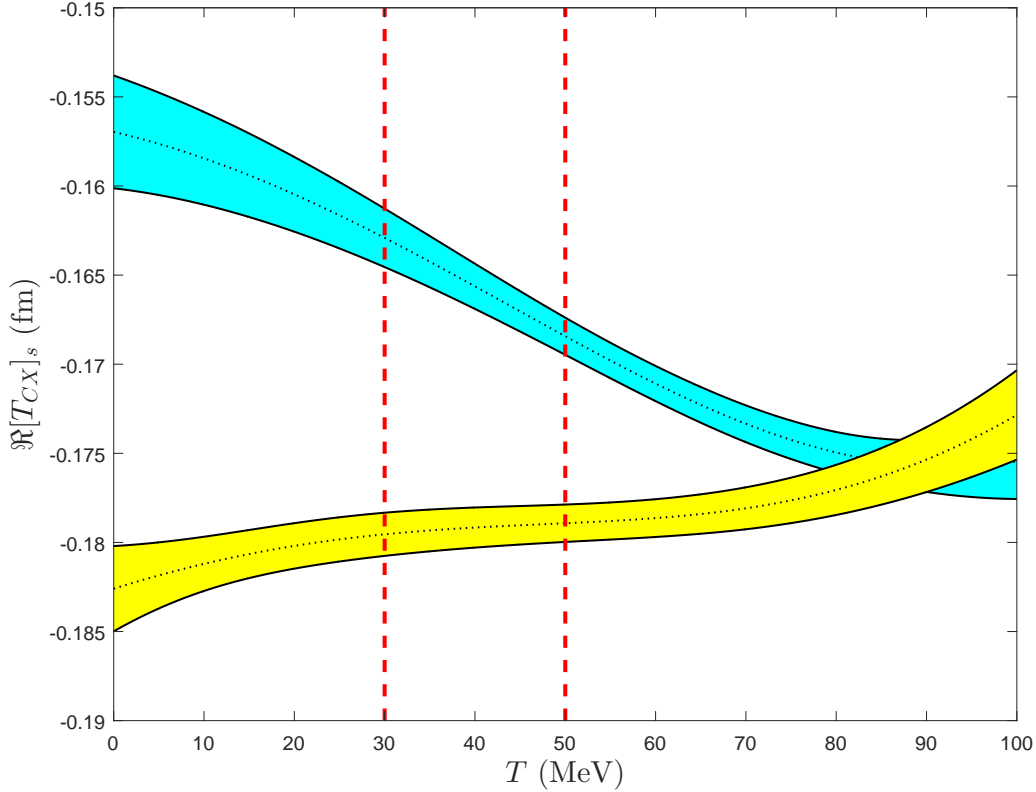}
\caption{\label{fig:ISB}Comparison of the real parts of the two $s$-wave $\pi^- p$ CX amplitudes, obtained via Eq.~(\ref{eq:EQ028}) from the results of this study. Blue band: the $\pi^- p$ CX amplitude is obtained via the 
triangle identity, e.g., see Eq.~(2) of Ref.~\cite{Matsinos2022a}, from the joint fit to the ES data using the ETH parameterisation of the $K$-matrix elements. Yellow band: the same quantity, obtained from the joint fit to 
the $\pi^+ p$ and $\pi^- p$ CX data. The two red dashed vertical straight lines mark the limits of the energy domain of Fig.~1 of Ref.~\cite{Gibbs1995}, which had been the first report on the violation of the isospin 
invariance in the $\pi N$ interaction at low energy.}
\vspace{0.5cm}
\end{center}
\end{figure}

The real parts of the two amplitudes, obtained from the results of the joint fits A and B, are shown in Fig.~\ref{fig:ISB}. It was over twenty-five years ago when such a comparison was made, in the first report on the 
violation of the isospin invariance in the $\pi N$ interaction at low energy \cite{Gibbs1995}, see Fig.~1 therein. Shown in that figure was the energy dependence of the two amplitudes in the energy domain between $T=30$ and 
$50$ MeV: their difference was found to be nearly constant, evaluated in Ref.~\cite{Gibbs1995} to $D \coloneqq (\Re [ T_{CX} ]_s)_{\rm B} - (\Re [ T_{CX} ]_s)_{\rm A} = -0.012(3)$ fm. Although the real part of the $s$-wave 
$\pi^- p$ CX amplitude from the joint fit A exhibits a more pronounced energy dependence (in comparison with the corresponding result of Ref.~\cite{Gibbs1995}), the estimate of this work for the difference $D$ (an average 
over the energy domain of Ref.~\cite{Gibbs1995}) is nearly unchanged: $D \approx -0.013$ fm. As Fig.~\ref{fig:ISB} of this work suggests, the energy dependence of the difference between the real parts of the two $s$-wave 
$\pi^- p$ CX amplitudes is pronounced in the low-energy region, decreasing with increasing energy and vanishing in the vicinity of about $80$ MeV.

Those who doubt the possibility of such large effects should take a better look at Fig.~14 of Ref.~\cite{Matsinos2017a}. Similar plots have been obtained from all analyses of the low-energy $\pi N$ data with the ETH model 
during the past two decades. (In comparison with the analyses using the parameterisations of this work, smaller uncertainties are generally obtained when fitting the ETH model to the same data; this is due to the constraint 
of crossing symmetry, which the ETH model fulfils.) It ought to be borne in mind that Fig.~14 of Ref.~\cite{Matsinos2017a} pertains to cross sections $\sigma$, whereas Fig.~\ref{fig:ISB} of this work to amplitudes $f$; 
generally speaking, $\sigma \sim \lvert f \rvert^2$.

\begin{figure}
\begin{center}
\includegraphics [width=15.5cm] {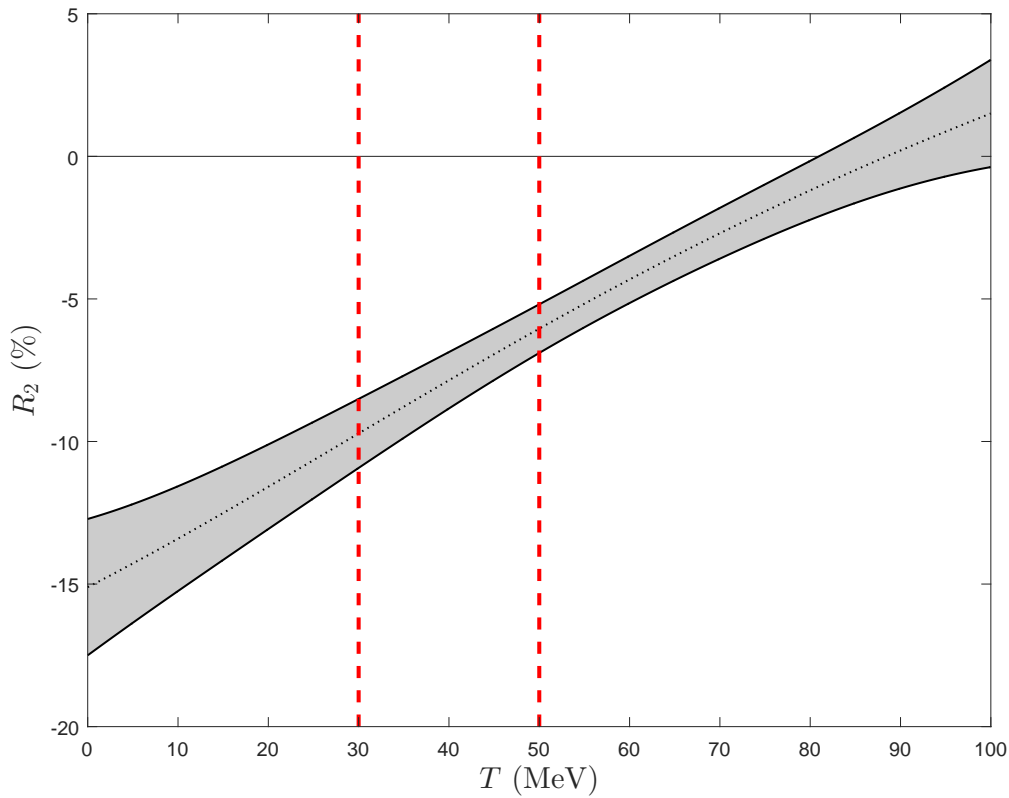}
\caption{\label{fig:R2}The indicator $R_2$ of Eq.~(3) of Ref.~\cite{Matsinos2022a}, which represents the symmetrised relative difference between the real parts of the two $s$-wave $\pi^- p$ CX amplitudes shown in Fig.~\ref{fig:ISB}.}
\vspace{0.5cm}
\end{center}
\end{figure}

One of the popular ways of quantifying the difference between the $\pi^- p$ CX scattering amplitudes, obtained from the joint fits A and B, employs the indicator $R_2$, see Eq.~(3) of Ref.~\cite{Matsinos2022a}, which 
represents the symmetrised relative difference between their real parts (evaluated separately for the $s$ waves, spin-flip and no-spin-flip $p$ waves, etc.). The energy dependence of the quantity $R_2$ for the $s$ wave is 
shown in Fig.~\ref{fig:R2}. The departure of this quantity from $0$ may be due to any of three reasons (or their combination) \cite{Matsinos2022a}:
\begin{itemize}
\item systematic effects in the absolute normalisation of the bulk of the low-energy $\pi N$ data,
\item sizeable residual contributions (i.e., at present not included) in the EM corrections (which are applied to the data in order that the hadronic quantities be extracted), and
\item the violation of the isospin invariance in the $\pi N$ interaction well beyond the $\chi$PT expectations \cite{Hoferichter2010}.
\end{itemize}

Another application of the results of this work can be found in Ref.~\cite{Matsinos2022b}.

\section{\label{sec:Conclusions}Conclusions}

The results of the application of four polynomial parameterisations of the $s$- and $p$-wave $K$-matrix elements (or of their reciprocal), suitable for the pion-nucleon ($\pi N$) interaction at low energy (pion laboratory 
kinetic energy $T \leq 100$ MeV), have been compared in this work. After the inclusion of the resonant contributions in two $p$ waves (see Section \ref{sec:Resonant}), of the $d$ and $f$ waves of the SAID phase-shift 
solution XP15 \cite{XP15}, and of the electromagnetic (EM) effects \cite{Oades2007,Gashi2001a,Gashi2001b}, fits were pursued (using these parameterisations) to a subset of the available $\pi N$ measurements, containing no 
outliers for any of the four modelling options of the $s$- and $p$-wave $K$-matrix elements. After comparing the final $\chi^2$ values of these fits, one of the parameterisations (the one detailed in Section \ref{sec:ETH2}) 
was not pursued further. The remaining three parameterisations are:
\begin{itemize}
\item the effective-range expansion, proposed by H{\"o}hler \cite{Hoehler1983}, see Section \ref{sec:Hoehler};
\item a simple polynomial parameterisation in $q^2$ (ELW parameterisation), where $q$ is the magnitude of the $3$-momentum vector in the centre-of-mass (CM) coordinate system, also proposed by H{\"o}hler \cite{Hoehler1983} 
and put in application in a 2004 paper by Ericson, Loiseau, and Wycech \cite{Ericson2004}, see Section \ref{sec:ELW}; and
\item a parameterisation in the pion CM kinetic energy $\epsilon$ (ETH parameterisation), put in application in a 1997 paper by Fettes and myself \cite{Fettes1997}, see Section \ref{sec:ETH1}.
\end{itemize}

These three parameterisations were applied to the initial low-energy $\pi N$ database, as it has been detailed in Ref.~\cite{Matsinos2017a}. After the application of the same procedure, see steps (1)-(5) of Section \ref{sec:Procedure}, 
the outliers were removed from the data, separately for the three parameterisations, as each method requires. Initial and final $\chi^2$ values from the application of these methods to the data are given in Tables 
\ref{tab:HOEHLER}, \ref{tab:ELW}, and \ref{tab:ETH}: the three methods achieve comparable descriptions of the input data. Noticeable is only a marginal difficulty of the forms, associated with the effective-range expansion, 
to account for the $p$-wave part of the two $\pi^- p$ (ES and CX) reactions.

The fitted values and uncertainties of the model parameters for the three parameterisations can be found in Tables \ref{tab:HOEHLERParameters}, \ref{tab:ELWParameters}, and \ref{tab:ETHParameters} for two types of fit:
\begin{itemize}
\item joint fit to the measurements of the two elastic-scattering (ES) reactions $\pi^\pm p \to \pi^\pm p$ (joint fit A) and
\item joint fit to the measurements of the $\pi^+ p$ reaction and of the $\pi^- p$ charge-exchange (CX) reaction $\pi^- p \to \pi^0 n$ (joint fit B).
\end{itemize}
In both cases, the isospin $I=3/2$ phases shifts are largely determined from the $\pi^+ p$ reaction, leaving the determination of the $I=1/2$ phase shifts to the corresponding $\pi^- p$ (ES or CX) reaction. The Hessian 
(covariance) matrices from these fits are given in Tables
\begin{itemize}
\item \ref{tab:HOEHLERHessian1} and \ref{tab:HOEHLERHessian2} for the effective-range expansion,
\item \ref{tab:ELWHessian1} and \ref{tab:ELWHessian2} for the ELW parameterisation, and
\item \ref{tab:ETHHessian1} and \ref{tab:ETHHessian2} for the ETH parameterisation.
\end{itemize}
These results are also uploaded as ancillary material, in the form of one Excel file; more precise information (i.e., more decimal places) about the fitted values and uncertainties of the model parameters can be found in 
that file.

From these results, reliable and (largely) data-driven (hence model-independent) predictions, accompanied by uncertainties which reflect the statistical and systematic fluctuation of the input data, can be obtained 
for the low-energy constants of the $\pi N$ interaction (scattering lengths/volumes and range parameters), for the $\pi N$ phase shifts, for the $K$-matrix elements, and for the partial-wave amplitudes in the (dominant at 
low energy) $s$ and $p$ waves. The hope is that the interested users will obtain their predictions (using the parameterisation of their choice) from the results of this work, which are based on the modern (meson-factory) 
$\pi N$ measurements, rather than seek the importation of the corresponding information from the outdated analyses of the Karlsruhe programme \cite{Hoehler1983}.

After the addition of the $d$- and $f$-wave contributions and the inclusion of the EM effects, corresponding predictions can be obtained for the observables of the three $\pi N$ reactions which are experimentally accessible 
at low energy, namely of the two ES processes and of the $\pi^- p$ CX reaction. Due to the non-fulfilment of the triangle identity, see Eq.~(2) of Ref.~\cite{Matsinos2022a}, by the scattering amplitudes of the three 
low-energy $\pi N$ reactions, the recommendation of this work is to make use of the results from the joint fit A when the objective is the generation of a prediction associated with any of the two ES processes, and from the 
joint fit B when the objective is the generation of a prediction involving the $\pi^- p$ CX reaction.

\begin{ack}
The figures in this paper were created with MATLAB$^{\textregistered}$~(The MathWorks, Inc., Natick, Massachusetts, United States).
\end{ack}

\clearpage
\vspace{0.5cm}
\begin{table}[h!]
{\bf \caption{\label{tab:HOEHLER}}}The $\chi^2$ values for steps (1)-(5) of Section \ref{sec:Procedure} for the effective-range expansion (Section \ref{sec:Hoehler}). The outliers were removed one at a time, starting from 
the initial DBs, as they have been detailed in Ref.~\cite{Matsinos2017a}. The application of the method resulted in the removal of $70$ degrees of freedom from the low-energy $\pi N$ DBs.
\vspace{0.25cm}
\begin{center}
\begin{tabular}{|l|c|c|c|c|}
\hline
DB & Remark & $\chi^2$ & NDF & p-value\\
\hline
\hline
$\pi^+ p$ & Initial fit & $925.26$ & $452$ & $8.85 \cdot 10^{-35}$\\
 & Final fit & $544.48$ & $414$ & $1.64 \cdot 10^{-5}$\\
\hline
$\pi^- p$ ES & Initial fit & $533.44$ & $334$ & $2.25 \cdot 10^{-11}$\\
 & Final fit & $380.74$ & $325$ & $1.79 \cdot 10^{-2}$\\
\hline
$\pi^- p$ CX & Initial fit & $400.15$ & $327$ & $3.50 \cdot 10^{-3}$\\
 & Final fit & $322.70$ & $322$ & $4.79 \cdot 10^{-1}$\\
\hline
ES & Initial fit & $909.39$ & $739$ & $1.68 \cdot 10^{-5}$\\
 & Final fit & $873.75$ & $721$ & $7.64 \cdot 10^{-5}$\\
\hline
$\pi^+ p$ and $\pi^- p$ CX & Initial fit & $851.01$ & $736$ & $2.04 \cdot 10^{-3}$\\
 & Final fit & $851.01$ & $736$ & $2.04 \cdot 10^{-3}$\\
\hline
\end{tabular}
\end{center}
\vspace{0.5cm}
\end{table}

\vspace{0.5cm}
\begin{table}[h!]
{\bf \caption{\label{tab:HOEHLERParameters}}}The fitted values and uncertainties of the model parameters for the effective-range expansion (Section \ref{sec:Hoehler}). The two sets of results correspond to two types of joint 
fits, namely to the ES DB, and to the $\pi^+ p$ and $\pi^- p$ CX DB. The fitted uncertainties have been corrected via the application of the Birge factor $\sqrt{\chi^2/{\rm NDF}}$, which takes account of the goodness of each 
fit \cite{Birge1932}. It ought to be borne in mind that the numerical results for the quantities $a^{3/2}_{1+}$, $b^{3/2}_{1+}$, $a^{1/2}_{1-}$, and $b^{1/2}_{1-}$ correspond to the \emph{background} contributions; they do 
not contain any effects from the resonant parts, detailed in Section \ref{sec:Resonant}.
\vspace{0.25cm}
\begin{center}
\begin{tabular}{|l|c|c|c|c|}
\hline
 & \multicolumn{2}{c|}{ES, joint fit A} & \multicolumn{2}{c|}{$\pi^+ p$ and $\pi^- p$ CX, joint fit B}\\
\hline
Parameter (unit) & Value & Uncertainty & Value & Uncertainty\\
\hline
\hline
$a^{3/2}_{0+}$ (GeV$^{-1}$) & $-0.531$ & $0.021$ & $-0.541$ & $0.020$\\
$b^{3/2}_{0+}$ (GeV$^{-1}$) & $67.9$ & $9.2$ & $65.7$ & $8.9$\\
$c^{3/2}_{0+}$ (GeV$^{-3}$) & $-1322$ & $273$ & $-1291$ & $275$\\
$a^{3/2}_{1+}$ (GeV$^{-3}$) & $38.5$ & $1.4$ & $36.3$ & $1.2$\\
$b^{3/2}_{1+}$ (GeV) & $0.397$ & $0.055$ & $0.326$ & $0.053$\\
$a^{3/2}_{1-}$ (GeV$^{-3}$) & $-18.1$ & $2.1$ & $-14.3$ & $1.5$\\
$b^{3/2}_{1-}$ (GeV) & $-2.43$ & $0.41$ & $-1.65$ & $0.44$\\
\hline
$a^{1/2}_{0+}$ (GeV$^{-1}$) & $1.184$ & $0.012$ & $1.427$ & $0.033$\\
$b^{1/2}_{0+}$ (GeV$^{-1}$) & $12.9$ & $1.7$ & $12.9$ & $3.1$\\
$c^{1/2}_{0+}$ (GeV$^{-3}$) & $-173$ & $83$ & $117$ & $148$\\
$a^{1/2}_{1+}$ (GeV$^{-3}$) & $-5.09$ & $0.80$ & $-24.4$ & $5.4$\\
$b^{1/2}_{1+}$ (GeV) & $1.7$ & $1.5$ & $-2.91$ & $0.67$\\
$a^{1/2}_{1-}$ (GeV$^{-3}$) & $-65$ & $11$ & $-35.1$ & $8.3$\\
$b^{1/2}_{1-}$ (GeV) & $-2.77$ & $0.17$ & $-2.13$ & $0.43$\\
\hline
\end{tabular}
\end{center}
\vspace{0.5cm}
\end{table}

\begin{sidewaystable}[h]
{\bf \caption{\label{tab:HOEHLERHessian1}}}The Hessian (covariance) matrix from the joint fit to the low-energy $\pi N$ ES data (joint fit A) using the effective-range expansion (Section \ref{sec:Hoehler}). The rows (top to 
bottom) and the columns (left to right) of this table follow the order in which the fourteen model parameters are listed in Tables \ref{tab:HOEHLERParameters}, \ref{tab:ELWParameters}, and \ref{tab:ETHParameters}, i.e., 
first the isospin $I=3/2$ parameters $a^{3/2}_{0+}$, \dots $b^{3/2}_{1-}$, followed by the $I=1/2$ parameters $a^{1/2}_{0+}$, \dots $b^{1/2}_{1-}$.
\vspace{0.2cm}
\tiny
\begin{center}
\begin{tabular}{cccccccccccccc}
$1.000000$ & $0.976324$ & $-0.935777$ & $0.087047$ & $0.079544$ & $-0.431295$ & $-0.352238$ & $-0.901686$ & $-0.428652$ & $0.231559$ & $-0.170976$ & $-0.171465$ & $0.213624$ & $0.164906$\\
$0.976324$ & $1.000000$ & $-0.988196$ & $0.127173$ & $0.114787$ & $-0.402072$ & $-0.328756$ & $-0.880206$ & $-0.478174$ & $0.293845$ & $-0.234053$ & $-0.233830$ & $0.215012$ & $0.168793$\\
$-0.935777$ & $-0.988196$ & $1.000000$ & $-0.152604$ & $-0.140603$ & $0.376691$ & $0.309090$ & $0.843884$ & $0.499105$ & $-0.332458$ & $0.269916$ & $0.274928$ & $-0.203054$ & $-0.154043$\\
$0.087047$ & $0.127173$ & $-0.152604$ & $1.000000$ & $0.947207$ & $-0.278293$ & $-0.218991$ & $-0.081279$ & $-0.048825$ & $0.022961$ & $-0.277876$ & $-0.311300$ & $-0.003371$ & $0.004186$\\
$0.079544$ & $0.114787$ & $-0.140603$ & $0.947207$ & $1.000000$ & $-0.241753$ & $-0.190893$ & $-0.072789$ & $-0.032517$ & $0.023798$ & $-0.235940$ & $-0.291666$ & $0.013679$ & $0.045861$\\
$-0.431295$ & $-0.402072$ & $0.376691$ & $-0.278293$ & $-0.241753$ & $1.000000$ & $0.926523$ & $0.388812$ & $0.129962$ & $-0.029990$ & $0.073430$ & $0.082236$ & $-0.312705$ & $-0.294127$\\
$-0.352238$ & $-0.328756$ & $0.309090$ & $-0.218991$ & $-0.190893$ & $0.926523$ & $1.000000$ & $0.317119$ & $0.099826$ & $-0.017359$ & $0.046693$ & $0.054799$ & $-0.317817$ & $-0.370025$\\
$-0.901686$ & $-0.880206$ & $0.843884$ & $-0.081279$ & $-0.072789$ & $0.388812$ & $0.317119$ & $1.000000$ & $0.522669$ & $-0.300282$ & $0.147984$ & $0.148119$ & $-0.183749$ & $-0.141570$\\
$-0.428652$ & $-0.478174$ & $0.499105$ & $-0.048825$ & $-0.032517$ & $0.129962$ & $0.099826$ & $0.522669$ & $1.000000$ & $-0.933358$ & $0.422276$ & $0.387271$ & $-0.526331$ & $-0.402442$\\
$0.231559$ & $0.293845$ & $-0.332458$ & $0.022961$ & $0.023798$ & $-0.029990$ & $-0.017359$ & $-0.300282$ & $-0.933358$ & $1.000000$ & $-0.357911$ & $-0.318774$ & $0.500308$ & $0.410307$\\
$-0.170976$ & $-0.234053$ & $0.269916$ & $-0.277876$ & $-0.235940$ & $0.073430$ & $0.046693$ & $0.147984$ & $0.422276$ & $-0.357911$ & $1.000000$ & $0.978978$ & $-0.517032$ & $-0.308298$\\
$-0.171465$ & $-0.233830$ & $0.274928$ & $-0.311300$ & $-0.291666$ & $0.082236$ & $0.054799$ & $0.148119$ & $0.387271$ & $-0.318774$ & $0.978978$ & $1.000000$ & $-0.488149$ & $-0.298068$\\
$0.213624$ & $0.215012$ & $-0.203054$ & $-0.003371$ & $0.013679$ & $-0.312705$ & $-0.317817$ & $-0.183749$ & $-0.526331$ & $0.500308$ & $-0.517032$ & $-0.488149$ & $1.000000$ & $0.907949$\\
$0.164906$ & $0.168793$ & $-0.154043$ & $0.004186$ & $0.045861$ & $-0.294127$ & $-0.370025$ & $-0.141570$ & $-0.402442$ & $0.410307$ & $-0.308298$ & $-0.298068$ & $0.907949$ & $1.000000$\\
\end{tabular}
\end{center}
\end{sidewaystable}

\begin{sidewaystable}[h]
{\bf \caption{\label{tab:HOEHLERHessian2}}}The Hessian (covariance) matrix from the joint fit to the low-energy $\pi^+ p$ and $\pi^- p$ CX data (joint fit B) using the effective-range expansion (Section \ref{sec:Hoehler}). 
The rows (top to bottom) and the columns (left to right) of this table follow the order in which the fourteen model parameters are listed in Tables \ref{tab:HOEHLERParameters}, \ref{tab:ELWParameters}, and \ref{tab:ETHParameters}, 
i.e., first the isospin $I=3/2$ parameters $a^{3/2}_{0+}$, \dots $b^{3/2}_{1-}$, followed by the $I=1/2$ parameters $a^{1/2}_{0+}$, \dots $b^{1/2}_{1-}$.
\vspace{0.2cm}
\tiny
\begin{center}
\begin{tabular}{cccccccccccccc}
$1.000000$ & $0.966445$ & $-0.911533$ & $-0.072053$ & $-0.054644$ & $-0.354642$ & $-0.297024$ & $0.656472$ & $0.319723$ & $-0.111687$ & $-0.039514$ & $-0.012191$ & $-0.068771$ & $-0.038438$\\
$0.966445$ & $1.000000$ & $-0.984013$ & $-0.056779$ & $-0.047226$ & $-0.280213$ & $-0.233652$ & $0.644672$ & $0.392211$ & $-0.188743$ & $-0.031980$ & $-0.015117$ & $-0.049179$ & $-0.021354$\\
$-0.911533$ & $-0.984013$ & $1.000000$ & $0.054040$ & $0.054332$ & $0.228168$ & $0.189889$ & $-0.615940$ & $-0.429223$ & $0.247824$ & $0.031128$ & $0.022278$ & $0.036230$ & $0.011860$\\
$-0.072053$ & $-0.056779$ & $0.054040$ & $1.000000$ & $0.933623$ & $-0.206033$ & $-0.176043$ & $-0.055610$ & $-0.053453$ & $0.067982$ & $0.353419$ & $0.229523$ & $-0.056220$ & $-0.046172$\\
$-0.054644$ & $-0.047226$ & $0.054332$ & $0.933623$ & $1.000000$ & $-0.176156$ & $-0.152145$ & $-0.047204$ & $-0.068504$ & $0.101099$ & $0.340584$ & $0.292984$ & $-0.041241$ & $-0.037761$\\
$-0.354642$ & $-0.280213$ & $0.228168$ & $-0.206033$ & $-0.176156$ & $1.000000$ & $0.940809$ & $-0.225371$ & $-0.045492$ & $-0.024085$ & $-0.057699$ & $-0.039274$ & $0.292271$ & $0.267609$\\
$-0.297024$ & $-0.233652$ & $0.189889$ & $-0.176043$ & $-0.152145$ & $0.940809$ & $1.000000$ & $-0.189596$ & $-0.041530$ & $-0.016106$ & $-0.047535$ & $-0.034312$ & $0.288866$ & $0.318157$\\
$0.656472$ & $0.644672$ & $-0.615940$ & $-0.055610$ & $-0.047204$ & $-0.225371$ & $-0.189596$ & $1.000000$ & $0.730572$ & $-0.474745$ & $-0.011155$ & $-0.007901$ & $-0.048091$ & $-0.034076$\\
$0.319723$ & $0.392211$ & $-0.429223$ & $-0.053453$ & $-0.068504$ & $-0.045492$ & $-0.041530$ & $0.730572$ & $1.000000$ & $-0.913965$ & $0.007135$ & $-0.016255$ & $0.186211$ & $0.144373$\\
$-0.111687$ & $-0.188743$ & $0.247824$ & $0.067982$ & $0.101099$ & $-0.024085$ & $-0.016106$ & $-0.474745$ & $-0.913965$ & $1.000000$ & $-0.030159$ & $-0.023978$ & $-0.219013$ & $-0.174465$\\
$-0.039514$ & $-0.031980$ & $0.031128$ & $0.353419$ & $0.340584$ & $-0.057699$ & $-0.047535$ & $-0.011155$ & $0.007135$ & $-0.030159$ & $1.000000$ & $0.902480$ & $-0.492791$ & $-0.422560$\\
$-0.012191$ & $-0.015117$ & $0.022278$ & $0.229523$ & $0.292984$ & $-0.039274$ & $-0.034312$ & $-0.007901$ & $-0.016255$ & $-0.023978$ & $0.902480$ & $1.000000$ & $-0.365055$ & $-0.339512$\\
$-0.068771$ & $-0.049179$ & $0.036230$ & $-0.056220$ & $-0.041241$ & $0.292271$ & $0.288866$ & $-0.048091$ & $0.186211$ & $-0.219013$ & $-0.492791$ & $-0.365055$ & $1.000000$ & $0.933317$\\
$-0.038438$ & $-0.021354$ & $0.011860$ & $-0.046172$ & $-0.037761$ & $0.267609$ & $0.318157$ & $-0.034076$ & $0.144373$ & $-0.174465$ & $-0.422560$ & $-0.339512$ & $0.933317$ & $1.000000$\\
\end{tabular}
\end{center}
\end{sidewaystable}

\vspace{0.5cm}
\begin{table}[h!]
{\bf \caption{\label{tab:ELW}}}The equivalent of Table \ref{tab:HOEHLER} for the ELW parameterisation (Section \ref{sec:ELW}). The application of the method resulted in the removal of $49$ degrees of freedom from the 
low-energy $\pi N$ DBs.
\vspace{0.25cm}
\begin{center}
\begin{tabular}{|l|c|c|c|c|}
\hline
DB & Remark & $\chi^2$ & NDF & p-value\\
\hline
\hline
$\pi^+ p$ & Initial fit & $929.51$ & $452$ & $2.95 \cdot 10^{-35}$\\
 & Final fit & $550.92$ & $414$ & $7.16 \cdot 10^{-6}$\\
\hline
$\pi^- p$ ES & Initial fit & $522.15$ & $334$ & $1.89 \cdot 10^{-10}$\\
 & Final fit & $364.89$ & $327$ & $7.30 \cdot 10^{-2}$\\
\hline
$\pi^- p$ CX & Initial fit & $378.75$ & $327$ & $2.55 \cdot 10^{-2}$\\
 & Final fit & $312.33$ & $323$ & $6.55 \cdot 10^{-1}$\\
\hline
ES & Initial fit & $905.44$ & $741$ & $3.06 \cdot 10^{-5}$\\
 & Final fit & $905.44$ & $741$ & $3.06 \cdot 10^{-5}$\\
\hline
$\pi^+ p$ and $\pi^- p$ CX & Initial fit & $846.69$ & $737$ & $3.04 \cdot 10^{-3}$\\
 & Final fit & $846.69$ & $737$ & $3.04 \cdot 10^{-3}$\\
\hline
\end{tabular}
\end{center}
\vspace{0.5cm}
\end{table}

\vspace{0.5cm}
\begin{table}[h!]
{\bf \caption{\label{tab:ELWParameters}}}The equivalent of Table \ref{tab:HOEHLERParameters} for the ELW parameterisation (Section \ref{sec:ELW}). 
\vspace{0.25cm}
\begin{center}
\begin{tabular}{|l|c|c|c|c|}
\hline
 & \multicolumn{2}{c|}{ES, joint fit A} & \multicolumn{2}{c|}{$\pi^+ p$ and $\pi^- p$ CX, joint fit B}\\
\hline
Parameter (unit) & Value & Uncertainty & Value & Uncertainty\\
\hline
\hline
$a^{3/2}_{0+}$ (GeV$^{-1}$) & $-0.473$ & $0.036$ & $-0.467$ & $0.040$\\
$b^{3/2}_{0+}$ (GeV$^{-3}$) & $-33.8$ & $5.1$ & $-35.5$ & $6.0$\\
$c^{3/2}_{0+}$ (GeV$^{-5}$) & $520$ & $164$ & $581$ & $203$\\
$a^{3/2}_{1+}$ (GeV$^{-3}$) & $35.21$ & $0.89$ & $35.13$ & $0.91$\\
$b^{3/2}_{1+}$ (GeV$^{-5}$) & $-282$ & $50$ & $-282$ & $52$\\
$a^{3/2}_{1-}$ (GeV$^{-3}$) & $-13.57$ & $0.91$ & $-13.38$ & $0.90$\\
$b^{3/2}_{1-}$ (GeV$^{-5}$) & $191$ & $49$ & $182$ & $48$\\
\hline
$a^{1/2}_{0+}$ (GeV$^{-1}$) & $1.156$ & $0.019$ & $1.495$ & $0.046$\\
$b^{1/2}_{0+}$ (GeV$^{-3}$) & $-9.3$ & $2.8$ & $-37.3$ & $7.4$\\
$c^{1/2}_{0+}$ (GeV$^{-5}$) & $36$ & $106$ & $613$ & $268$\\
$a^{1/2}_{1+}$ (GeV$^{-3}$) & $-5.93$ & $0.96$ & $-20.8$ & $1.9$\\
$b^{1/2}_{1+}$ (GeV$^{-5}$) & $-21$ & $48$ & $579$ & $112$\\
$a^{1/2}_{1-}$ (GeV$^{-3}$) & $-30.7$ & $1.2$ & $-26.3$ & $2.5$\\
$b^{1/2}_{1-}$ (GeV$^{-5}$) & $805$ & $62$ & $622$ & $135$\\
\hline
\end{tabular}
\end{center}
\vspace{0.5cm}
\end{table}

\begin{sidewaystable}[h]
{\bf \caption{\label{tab:ELWHessian1}}}The equivalent of Table \ref{tab:HOEHLERHessian1} for the ELW parameterisation (Section \ref{sec:ELW}).
\vspace{0.2cm}
\tiny
\begin{center}
\begin{tabular}{cccccccccccccc}
$1.000000$ & $-0.964018$ & $0.905086$ & $0.069548$ & $-0.057704$ & $-0.338115$ & $0.303394$ & $-0.963298$ & $0.821543$ & $-0.658151$ & $-0.088731$ & $0.084859$ & $0.045536$ & $-0.044591$\\
$-0.964018$ & $1.000000$ & $-0.981560$ & $-0.140677$ & $0.120235$ & $0.269253$ & $-0.241994$ & $0.927570$ & $-0.835033$ & $0.695574$ & $0.138688$ & $-0.136963$ & $0.005333$ & $-0.006565$\\
$0.905086$ & $-0.981560$ & $1.000000$ & $0.175525$ & $-0.156733$ & $-0.223234$ & $0.201435$ & $-0.870547$ & $0.811205$ & $-0.696132$ & $-0.167027$ & $0.177195$ & $-0.036279$ & $0.043995$\\
$0.069548$ & $-0.140677$ & $0.175525$ & $1.000000$ & $-0.952154$ & $-0.158269$ & $0.138605$ & $-0.068185$ & $0.153118$ & $-0.164789$ & $-0.416089$ & $0.430085$ & $0.123359$ & $-0.131319$\\
$-0.057704$ & $0.120235$ & $-0.156733$ & $-0.952154$ & $1.000000$ & $0.138895$ & $-0.122212$ & $0.055800$ & $-0.126397$ & $0.147065$ & $0.382100$ & $-0.424716$ & $-0.117107$ & $0.140273$\\
$-0.338115$ & $0.269253$ & $-0.223234$ & $-0.158269$ & $0.138895$ & $1.000000$ & $-0.954815$ & $0.327315$ & $-0.256867$ & $0.189114$ & $0.066346$ & $-0.064171$ & $-0.386971$ & $0.396202$\\
$0.303394$ & $-0.241994$ & $0.201435$ & $0.138605$ & $-0.122212$ & $-0.954815$ & $1.000000$ & $-0.293693$ & $0.231040$ & $-0.170846$ & $-0.057509$ & $0.055328$ & $0.369560$ & $-0.413615$\\
$-0.963298$ & $0.927570$ & $-0.870547$ & $-0.068185$ & $0.055800$ & $0.327315$ & $-0.293693$ & $1.000000$ & $-0.861931$ & $0.693620$ & $0.083007$ & $-0.078715$ & $-0.044600$ & $0.043963$\\
$0.821543$ & $-0.835033$ & $0.811205$ & $0.153118$ & $-0.126397$ & $-0.256867$ & $0.231040$ & $-0.861931$ & $1.000000$ & $-0.948345$ & $-0.237862$ & $0.210450$ & $0.258086$ & $-0.249575$\\
$-0.658151$ & $0.695574$ & $-0.696132$ & $-0.164789$ & $0.147065$ & $0.189114$ & $-0.170846$ & $0.693620$ & $-0.948345$ & $1.000000$ & $0.250386$ & $-0.215481$ & $-0.313989$ & $0.317019$\\
$-0.088731$ & $0.138688$ & $-0.167027$ & $-0.416089$ & $0.382100$ & $0.066346$ & $-0.057509$ & $0.083007$ & $-0.237862$ & $0.250386$ & $1.000000$ & $-0.970750$ & $-0.578341$ & $0.527090$\\
$0.084859$ & $-0.136963$ & $0.177195$ & $0.430085$ & $-0.424716$ & $-0.064171$ & $0.055328$ & $-0.078715$ & $0.210450$ & $-0.215481$ & $-0.970750$ & $1.000000$ & $0.521613$ & $-0.479720$\\
$0.045536$ & $0.005333$ & $-0.036279$ & $0.123359$ & $-0.117107$ & $-0.386971$ & $0.369560$ & $-0.044600$ & $0.258086$ & $-0.313989$ & $-0.578341$ & $0.521613$ & $1.000000$ & $-0.971080$\\
$-0.044591$ & $-0.006565$ & $0.043995$ & $-0.131319$ & $0.140273$ & $0.396202$ & $-0.413615$ & $0.043963$ & $-0.249575$ & $0.317019$ & $0.527090$ & $-0.479720$ & $-0.971080$ & $1.000000$\\
\end{tabular}
\end{center}
\end{sidewaystable}

\begin{sidewaystable}[h]
{\bf \caption{\label{tab:ELWHessian2}}}The equivalent of Table \ref{tab:HOEHLERHessian2} for the ELW parameterisation (Section \ref{sec:ELW}).
\vspace{0.2cm}
\tiny
\begin{center}
\begin{tabular}{cccccccccccccc}
$1.000000$ & $-0.966784$ & $0.915566$ & $-0.082884$ & $0.083663$ & $-0.263533$ & $0.236008$ & $0.857756$ & $-0.779032$ & $0.678902$ & $-0.045480$ & $0.055830$ & $-0.104798$ & $0.090887$\\
$-0.966784$ & $1.000000$ & $-0.983982$ & $0.063883$ & $-0.072916$ & $0.177883$ & $-0.159272$ & $-0.827147$ & $0.801557$ & $-0.724283$ & $0.030158$ & $-0.048817$ & $0.082319$ & $-0.070700$\\
$0.915566$ & $-0.983982$ & $1.000000$ & $-0.062869$ & $0.081428$ & $-0.130567$ & $0.117272$ & $0.781929$ & $-0.785307$ & $0.730486$ & $-0.029496$ & $0.053774$ & $-0.067621$ & $0.058502$\\
$-0.082884$ & $0.063883$ & $-0.062869$ & $1.000000$ & $-0.947178$ & $-0.173196$ & $0.152824$ & $-0.075372$ & $0.063353$ & $-0.064182$ & $0.489121$ & $-0.429750$ & $-0.079503$ & $0.071587$\\
$0.083663$ & $-0.072916$ & $0.081428$ & $-0.947178$ & $1.000000$ & $0.148703$ & $-0.131791$ & $0.075617$ & $-0.071145$ & $0.081682$ & $-0.452251$ & $0.442141$ & $0.066061$ & $-0.061321$\\
$-0.263533$ & $0.177883$ & $-0.130567$ & $-0.173196$ & $0.148703$ & $1.000000$ & $-0.955278$ & $-0.227806$ & $0.145912$ & $-0.099261$ & $-0.075969$ & $0.060788$ & $0.352744$ & $-0.338043$\\
$0.236008$ & $-0.159272$ & $0.117272$ & $0.152824$ & $-0.131791$ & $-0.955278$ & $1.000000$ & $0.204079$ & $-0.130683$ & $0.089106$ & $0.066799$ & $-0.053650$ & $-0.337355$ & $0.354673$\\
$0.857756$ & $-0.827147$ & $0.781929$ & $-0.075372$ & $0.075617$ & $-0.227806$ & $0.204079$ & $1.000000$ & $-0.925581$ & $0.820179$ & $-0.025346$ & $0.040998$ & $-0.090306$ & $0.079591$\\
$-0.779032$ & $0.801557$ & $-0.785307$ & $0.063353$ & $-0.071145$ & $0.145912$ & $-0.130683$ & $-0.925581$ & $1.000000$ & $-0.967003$ & $-0.030965$ & $0.003560$ & $-0.043281$ & $0.042984$\\
$0.678902$ & $-0.724283$ & $0.730486$ & $-0.064182$ & $0.081682$ & $-0.099261$ & $0.089106$ & $0.820179$ & $-0.967003$ & $1.000000$ & $0.056342$ & $-0.033865$ & $0.098748$ & $-0.095514$\\
$-0.045480$ & $0.030158$ & $-0.029496$ & $0.489121$ & $-0.452251$ & $-0.075969$ & $0.066799$ & $-0.025346$ & $-0.030965$ & $0.056342$ & $1.000000$ & $-0.947935$ & $-0.443573$ & $0.413324$\\
$0.055830$ & $-0.048817$ & $0.053774$ & $-0.429750$ & $0.442141$ & $0.060788$ & $-0.053650$ & $0.040998$ & $0.003560$ & $-0.033865$ & $-0.947935$ & $1.000000$ & $0.378402$ & $-0.361423$\\
$-0.104798$ & $0.082319$ & $-0.067621$ & $-0.079503$ & $0.066061$ & $0.352744$ & $-0.337355$ & $-0.090306$ & $-0.043281$ & $0.098748$ & $-0.443573$ & $0.378402$ & $1.000000$ & $-0.970382$\\
$0.090887$ & $-0.070700$ & $0.058502$ & $0.071587$ & $-0.061321$ & $-0.338043$ & $0.354673$ & $0.079591$ & $0.042984$ & $-0.095514$ & $0.413324$ & $-0.361423$ & $-0.970382$ & $1.000000$\\
\end{tabular}
\end{center}
\end{sidewaystable}

\vspace{0.5cm}
\begin{table}[h!]
{\bf \caption{\label{tab:ETH}}}The equivalent of Table \ref{tab:HOEHLER} for the ETH parameterisation (Section \ref{sec:ETH1}). The application of the method resulted in the removal of $50$ degrees of freedom from the 
low-energy $\pi N$ DBs.
\vspace{0.25cm}
\begin{center}
\begin{tabular}{|l|c|c|c|c|}
\hline
DB & Remark & $\chi^2$ & NDF & p-value\\
\hline
\hline
$\pi^+ p$ & Initial fit & $926.87$ & $452$ & $5.85 \cdot 10^{-35}$\\
 & Final fit & $546.84$ & $414$ & $1.22 \cdot 10^{-5}$\\
\hline
$\pi^- p$ ES & Initial fit & $522.98$ & $334$ & $1.62 \cdot 10^{-10}$\\
 & Final fit & $367.76$ & $326$ & $5.52 \cdot 10^{-2}$\\
\hline
$\pi^- p$ CX & Initial fit & $390.90$ & $327$ & $8.70 \cdot 10^{-3}$\\
 & Final fit & $319.27$ & $323$ & $5.48 \cdot 10^{-1}$\\
\hline
ES & Initial fit & $903.72$ & $740$ & $3.25 \cdot 10^{-5}$\\
 & Final fit & $903.72$ & $740$ & $3.25 \cdot 10^{-5}$\\
\hline
$\pi^+ p$ and $\pi^- p$ CX & Initial fit & $850.14$ & $737$ & $2.36 \cdot 10^{-3}$\\
 & Final fit & $850.14$ & $737$ & $2.36 \cdot 10^{-3}$\\
\hline
\end{tabular}
\end{center}
\vspace{0.5cm}
\end{table}

\vspace{0.5cm}
\begin{table}[h!]
{\bf \caption{\label{tab:ETHParameters}}}The equivalent of Table \ref{tab:HOEHLERParameters} for the ETH parameterisation (Section \ref{sec:ETH1}). 
\vspace{0.25cm}
\begin{center}
\begin{tabular}{|l|c|c|c|c|}
\hline
 & \multicolumn{2}{c|}{ES, joint fit A} & \multicolumn{2}{c|}{$\pi^+ p$ and $\pi^- p$ CX, joint fit B}\\
\hline
Parameter (unit) & Value & Uncertainty & Value & Uncertainty\\
\hline
\hline
$a^{3/2}_{0+}$ (GeV$^{-1}$) & $-0.512$ & $0.022$ & $-0.520$ & $0.022$\\
$b^{3/2}_{0+}$ & $24.6$ & $3.6$ & $23.7$ & $3.6$\\
$c^{3/2}_{0+}$ (GeV$^{-1}$) & $-160$ & $36$ & $-153$ & $38$\\
$a^{3/2}_{1+}$ (GeV$^{-2}$) & $10.23$ & $0.31$ & $10.20$ & $0.32$\\
$b^{3/2}_{1+}$ (GeV$^{-3}$) & $-4.2$ & $6.1$ & $-4.2$ & $6.4$\\
$a^{3/2}_{1-}$ (GeV$^{-2}$) & $-4.13$ & $0.32$ & $-4.02$ & $0.32$\\
$b^{3/2}_{1-}$ (GeV$^{-3}$) & $14.3$ & $6.0$ & $12.5$ & $5.9$\\
\hline
$a^{1/2}_{0+}$ (GeV$^{-1}$) & $1.175$ & $0.012$ & $1.443$ & $0.035$\\
$b^{1/2}_{0+}$ & $3.04$ & $0.56$ & $3.7$ & $1.1$\\
$c^{1/2}_{0+}$ (GeV$^{-1}$) & $-2.1$ & $9.4$ & $29$ & $17$\\
$a^{1/2}_{1+}$ (GeV$^{-2}$) & $-1.42$ & $0.37$ & $-6.51$ & $0.68$\\
$b^{1/2}_{1+}$ (GeV$^{-3}$) & $-12.1$ & $6.3$ & $56$ & $14$\\
$a^{1/2}_{1-}$ (GeV$^{-2}$) & $-10.10$ & $0.47$ & $-8.15$ & $0.91$\\
$b^{1/2}_{1-}$ (GeV$^{-3}$) & $88.6$ & $8.0$ & $59$ & $17$\\
\hline
\end{tabular}
\end{center}
\vspace{0.5cm}
\end{table}

\begin{sidewaystable}[h]
{\bf \caption{\label{tab:ETHHessian1}}}The equivalent of Table \ref{tab:HOEHLERHessian1} for the ETH parameterisation (Section \ref{sec:ETH1}).
\vspace{0.2cm}
\tiny
\begin{center}
\begin{tabular}{cccccccccccccc}
$1.000000$ & $0.980130$ & $-0.943569$ & $0.001246$ & $-0.007922$ & $-0.387625$ & $0.346323$ & $-0.914348$ & $-0.459295$ & $0.221765$ & $-0.086781$ & $0.081397$ & $0.109904$ & $-0.103723$\\
$0.980130$ & $1.000000$ & $-0.989198$ & $0.051840$ & $-0.051454$ & $-0.348173$ & $0.311265$ & $-0.896090$ & $-0.493786$ & $0.267040$ & $-0.116674$ & $0.111944$ & $0.074921$ & $-0.069246$\\
$-0.943569$ & $-0.989198$ & $1.000000$ & $-0.085333$ & $0.083359$ & $0.316070$ & $-0.283057$ & $0.862872$ & $0.508065$ & $-0.297377$ & $0.138681$ & $-0.139943$ & $-0.048195$ & $0.039633$\\
$0.001246$ & $0.051840$ & $-0.085333$ & $1.000000$ & $-0.959100$ & $-0.173083$ & $0.153592$ & $-0.003697$ & $-0.131207$ & $0.127314$ & $-0.399107$ & $0.420485$ & $0.127536$ & $-0.136714$\\
$-0.007922$ & $-0.051454$ & $0.083359$ & $-0.959100$ & $1.000000$ & $0.157715$ & $-0.140835$ & $0.008479$ & $0.113768$ & $-0.125188$ & $0.364422$ & $-0.409202$ & $-0.119907$ & $0.141719$\\
$-0.387625$ & $-0.348173$ & $0.316070$ & $-0.173083$ & $0.157715$ & $1.000000$ & $-0.960384$ & $0.355504$ & $0.164396$ & $-0.063520$ & $0.094914$ & $-0.093797$ & $-0.376628$ & $0.389320$\\
$0.346323$ & $0.311265$ & $-0.283057$ & $0.153592$ & $-0.140835$ & $-0.960384$ & $1.000000$ & $-0.317631$ & $-0.147801$ & $0.058274$ & $-0.083674$ & $0.082646$ & $0.360485$ & $-0.402289$\\
$-0.914348$ & $-0.896090$ & $0.862872$ & $-0.003697$ & $0.008479$ & $0.355504$ & $-0.317631$ & $1.000000$ & $0.553118$ & $-0.289910$ & $0.077212$ & $-0.072099$ & $-0.099797$ & $0.094624$\\
$-0.459295$ & $-0.493786$ & $0.508065$ & $-0.131207$ & $0.113768$ & $0.164396$ & $-0.147801$ & $0.553118$ & $1.000000$ & $-0.926014$ & $0.328272$ & $-0.288112$ & $-0.439163$ & $0.427054$\\
$0.221765$ & $0.267040$ & $-0.297377$ & $0.127314$ & $-0.125188$ & $-0.063520$ & $0.058274$ & $-0.289910$ & $-0.926014$ & $1.000000$ & $-0.296242$ & $0.252657$ & $0.429596$ & $-0.433576$\\
$-0.086781$ & $-0.116674$ & $0.138681$ & $-0.399107$ & $0.364422$ & $0.094914$ & $-0.083674$ & $0.077212$ & $0.328272$ & $-0.296242$ & $1.000000$ & $-0.975692$ & $-0.623902$ & $0.573013$\\
$0.081397$ & $0.111944$ & $-0.139943$ & $0.420485$ & $-0.409202$ & $-0.093797$ & $0.082646$ & $-0.072099$ & $-0.288112$ & $0.252657$ & $-0.975692$ & $1.000000$ & $0.568722$ & $-0.526709$\\
$0.109904$ & $0.074921$ & $-0.048195$ & $0.127536$ & $-0.119907$ & $-0.376628$ & $0.360485$ & $-0.099797$ & $-0.439163$ & $0.429596$ & $-0.623902$ & $0.568722$ & $1.000000$ & $-0.976009$\\
$-0.103723$ & $-0.069246$ & $0.039633$ & $-0.136714$ & $0.141719$ & $0.389320$ & $-0.402289$ & $0.094624$ & $0.427054$ & $-0.433576$ & $0.573013$ & $-0.526709$ & $-0.976009$ & $1.000000$\\
\end{tabular}
\end{center}
\end{sidewaystable}

\begin{sidewaystable}[h]
{\bf \caption{\label{tab:ETHHessian2}}}The equivalent of Table \ref{tab:HOEHLERHessian2} for the ETH parameterisation (Section \ref{sec:ETH1}).
\vspace{0.2cm}
\tiny
\begin{center}
\begin{tabular}{cccccccccccccc}
$1.000000$ & $0.973806$ & $-0.928364$ & $-0.094229$ & $0.082724$ & $-0.324813$ & $0.288868$ & $0.686049$ & $0.325758$ & $-0.099034$ & $-0.074914$ & $0.065261$ & $-0.108085$ & $0.092769$\\
$0.973806$ & $1.000000$ & $-0.986753$ & $-0.078843$ & $0.073559$ & $-0.259038$ & $0.229959$ & $0.676770$ & $0.391333$ & $-0.168760$ & $-0.063537$ & $0.059591$ & $-0.091476$ & $0.077152$\\
$-0.928364$ & $-0.986753$ & $1.000000$ & $0.075211$ & $-0.077377$ & $0.212432$ & $-0.188541$ & $-0.651958$ & $-0.427407$ & $0.223780$ & $0.060397$ & $-0.061145$ & $0.078242$ & $-0.065538$\\
$-0.094229$ & $-0.078843$ & $0.075211$ & $1.000000$ & $-0.954986$ & $-0.187731$ & $0.167651$ & $-0.068635$ & $-0.040457$ & $0.041913$ & $0.479872$ & $-0.430046$ & $-0.082073$ & $0.075918$\\
$0.082724$ & $0.073559$ & $-0.077377$ & $-0.954986$ & $1.000000$ & $0.165086$ & $-0.148290$ & $0.063269$ & $0.054847$ & $-0.070112$ & $-0.448304$ & $0.439990$ & $0.070385$ & $-0.066927$\\
$-0.324813$ & $-0.259038$ & $0.212432$ & $-0.187731$ & $0.165086$ & $1.000000$ & $-0.960589$ & $-0.212916$ & $-0.023680$ & $-0.044384$ & $-0.076605$ & $0.064608$ & $0.343508$ & $-0.334983$\\
$0.288868$ & $0.229959$ & $-0.188541$ & $0.167651$ & $-0.148290$ & $-0.960589$ & $1.000000$ & $0.189333$ & $0.020648$ & $0.039497$ & $0.068253$ & $-0.057906$ & $-0.330413$ & $0.349657$\\
$0.686049$ & $0.676770$ & $-0.651958$ & $-0.068635$ & $0.063269$ & $-0.212916$ & $0.189333$ & $1.000000$ & $0.713947$ & $-0.442585$ & $-0.026634$ & $0.025777$ & $-0.052911$ & $0.045282$\\
$0.325758$ & $0.391333$ & $-0.427407$ & $-0.040457$ & $0.054847$ & $-0.023680$ & $0.020648$ & $0.713947$ & $1.000000$ & $-0.913692$ & $0.069617$ & $-0.061803$ & $0.194782$ & $-0.183167$\\
$-0.099034$ & $-0.168760$ & $0.223780$ & $0.041913$ & $-0.070112$ & $-0.044384$ & $0.039497$ & $-0.442585$ & $-0.913692$ & $1.000000$ & $-0.092283$ & $0.098725$ & $-0.222517$ & $0.211100$\\
$-0.074914$ & $-0.063537$ & $0.060397$ & $0.479872$ & $-0.448304$ & $-0.076605$ & $0.068253$ & $-0.026634$ & $0.069617$ & $-0.092283$ & $1.000000$ & $-0.958072$ & $-0.460480$ & $0.432055$\\
$0.065261$ & $0.059591$ & $-0.061145$ & $-0.430046$ & $0.439990$ & $0.064608$ & $-0.057906$ & $0.025777$ & $-0.061803$ & $0.098725$ & $-0.958072$ & $1.000000$ & $0.397914$ & $-0.381280$\\
$-0.108085$ & $-0.091476$ & $0.078242$ & $-0.082073$ & $0.070385$ & $0.343508$ & $-0.330413$ & $-0.052911$ & $0.194782$ & $-0.222517$ & $-0.460480$ & $0.397914$ & $1.000000$ & $-0.974810$\\
$0.092769$ & $0.077152$ & $-0.065538$ & $0.075918$ & $-0.066927$ & $-0.334983$ & $0.349657$ & $0.045282$ & $-0.183167$ & $0.211100$ & $0.432055$ & $-0.381280$ & $-0.974810$ & $1.000000$\\
\end{tabular}
\end{center}
\end{sidewaystable}

\clearpage
\newpage
\appendix
\section{\label{App:AppA}Contributions from the HBRs (see Section \ref{sec:Resonant}) to the scattering volumes and range parameters}

Equations (\ref{eq:EQ024},\ref{eq:EQ025}) can be directly used in the evaluations of $K^{3/2}_{1+}$ and $K^{1/2}_{1-}$, respectively. However, there may be occasions in which the explicit contributions from the HBRs to the 
scattering volumes $a^{3/2}_{1+}$ and $a^{1/2}_{1-}$, as well as to the range parameters $b^{3/2}_{1+}$ and $b^{1/2}_{1-}$, are needed. This appendix provides analytical expressions for these contributions in case of the 
ELW and ETH parameterisations.

In the ELW parameterisation, the contributions from the HBRs to the scattering volumes and range parameters are additive. For each of the contributions to $K^{3/2}_{1+}$, one obtains
\begin{equation} \label{eq:EQA01}
\frac{\Gamma_\Delta M_\Delta}{2 q_\Delta^3 (p_{0 \Delta} + m_p)} \frac{(p_0 + m_p) q^2}{W (M_\Delta-W)} = a^{3/2}_{1+} q^2 + b^{3/2}_{1+} q^4 + \mathcal{O}(q^6) \, \, \, \text{or}
\end{equation}
\begin{equation} \label{eq:EQA02}
\alpha_\Delta \frac{p_0 + m_p}{W (M_\Delta-W)} = a^{3/2}_{1+} + b^{3/2}_{1+} q^2 + \mathcal{O}(q^4) \, \, \, ,
\end{equation}
where
\begin{equation} \label{eq:EQA03}
\alpha_\Delta \coloneqq \frac{\Gamma_\Delta M_\Delta}{2 q_\Delta^3 (p_{0 \Delta} + m_p)}
\end{equation}
is a constant for each such resonance. Expanding the left-hand side of Eq.~(\ref{eq:EQA02}) in $q^2$, one finally obtains
\begin{align} \label{eq:EQA04}
\alpha_\Delta \Bigg( & \frac{2 m_p}{(m_p + m_c) (M_\Delta - m_p - m_c)}\nonumber\\
 & + \frac{1}{(m_p + m_c) (M_\Delta - m_p - m_c)^2} \left( \frac{M_\Delta - m_p - m_c}{2 m_p} - \frac{M_\Delta - 2 m_p - 2 m_c}{m_c} \right) q^2\nonumber\\
 & + \mathcal{O}(q^4) \Bigg) = a^{3/2}_{1+} + b^{3/2}_{1+} q^2 + \mathcal{O}(q^4) \, \, \, .
\end{align}
The identification of the contributions to $a^{3/2}_{1+}$ and $b^{3/2}_{1+}$ is straightforward.

The contributions to $a^{1/2}_{1-}$ and $b^{1/2}_{1-}$ can be obtained similarly for each of the two $P_{11}$ resonances. The final expressions are:
\begin{equation} \label{eq:EQA05}
a^{1/2}_{1-} = \frac{\alpha_N (2 m_p + m_c)^2}{2 m_p (m_p + m_c) (M_N - m_p - m_c)} \, \, \, \text{and}
\end{equation}
\begin{align} \label{eq:EQA06}
b^{1/2}_{1-} = & \frac{\alpha_N (2 m_p + m_c)}{4 m^2_p (m_p+m_c) (M_N - m_p - m_c)^2}\nonumber\\
 & \left( \frac{(m_p + m_c) (2 m_p + m_c)}{m_c} - \frac{m_c (M_N - m_p - m_c)}{2 m_p} \right)
\end{align}
where
\begin{equation} \label{eq:EQA07}
\alpha_N \coloneqq \frac{\Gamma_N M_N (p_{0 N} + m_p)}{2 q_N^3 (M_N + m_p)^2} \, \, \, .
\end{equation}

The relation between the scattering volumes and the range parameters of the ELW and ETH parameterisations can be obtained after equating the corresponding coefficients of the first two orders in $q^2$ in the expression:
\begin{equation} \label{eq:EQA08}
(a^I_{1\pm})_{\rm ELW} \, q^2 + (b^I_{1\pm})_{\rm ELW} \, q^4 + \mathcal{O}(q^6) = (a^I_{1\pm})_{\rm ETH} \, \epsilon + (b^I_{1\pm})_{\rm ETH} \, \epsilon^2 + \mathcal{O}(\epsilon^3) \, \, \, .
\end{equation}
Using Eq.~(\ref{eq:EQ019}) and retaining only the first two terms in the expansions, one obtains
\begin{equation} \label{eq:EQA09}
(a^I_{1\pm})_{\rm ETH} = 2 m_c (a^I_{1\pm})_{\rm ELW} \, \, \, \text{and}
\end{equation}
\begin{equation} \label{eq:EQA10}
(b^I_{1\pm})_{\rm ETH} = (a^I_{1\pm})_{\rm ELW} + 4 m^2_c (b^I_{1\pm})_{\rm ELW} \, \, \, .
\end{equation}

Table \ref{tab:Contributions} provides numerical results for the contributions from the HBRs (see Section \ref{sec:Resonant}). The determination of the contributions to $b^I_{1\pm}$ in case of the effective-range expansion 
(Section \ref{sec:Hoehler}) is trickier, as the background effects and those relating to the HBRs are not additive; on the other hand, the contributions to $a^I_{1\pm}$ for that parameterisation are identical to the ones 
extracted for the ELW parameterisation. It is straightforward to use these results; for instance, the background $a^{3/2}_{1+}$ for the ELW parameterisation from the fits to the ES data is equal to $35.21(89)$ GeV$^{-3}$, 
see Table \ref{tab:ELWParameters}. Therefore, the scattering volume $a^{3/2}_{1+}$, representing the properties of the $P_{33}$ channel in that parameterisation, is equal to $(35.21(89)+36.01)$ GeV$^{-3} = 71.22(89)$ GeV$^{-3}$ 
or $0.1936(24)~m^{-3}_c$. The result when using the effective-range expansion would be equal to $0.2026(38)~m^{-3}_c$, whereas that of the ETH parameterisation (converted into the form of the $a^{3/2}_{1+}$ result for the 
other two parameterisations) would be: $0.1975(31)~m^{-3}_c$.

\vspace{0.5cm}
\begin{table}[h!]
{\bf \caption{\label{tab:Contributions}}}The numerical results for the contributions from the HBRs (see Section \ref{sec:Resonant}). In case of the ELW parameterisation, the scattering volumes $a^{I}_{1\pm}$ are expressed in 
GeV$^{-3}$, whereas the range parameters $b^{I}_{1\pm}$ in GeV$^{-5}$. In case of the ETH parameterisation, the scattering volumes $a^{I}_{1\pm}$ are expressed in GeV$^{-2}$, whereas the range parameters $b^{I}_{1\pm}$ in 
GeV$^{-3}$.
\vspace{0.25cm}
\begin{center}
\begin{tabular}{|l|c|c|}
\hline
Physical quantity & ELW parameterisation & ETH parameterisation\\
\hline
\hline
$a^{3/2}_{1+}$ & $36.0130$ & $10.0527$\\
$b^{3/2}_{1+}$ & $833.0807$ & $100.9263$\\
$a^{3/2}_{1-}$ & $4.9616$ & $1.3850$\\
$b^{3/2}_{1-}$ & $58.9069$ & $9.5516$\\
\hline
\end{tabular}
\end{center}
\vspace{0.5cm}
\end{table}

\end{document}